\documentclass[conference]{IEEEtran}
\IEEEoverridecommandlockouts

\usepackage{geometry}
\def\BibTeX{{\rm B\kern-.05em{\sc i\kern-.025em b}\kern-.08em
    T\kern-.1667em\lower.7ex\hbox{E}\kern-.125emX}}

\usepackage{xcolor}
\usepackage{epsfig, amsmath, amssymb}
\usepackage{wrapfig,lipsum,booktabs}

\usepackage{epstopdf}
\usepackage{balance}
\usepackage{color}
\usepackage{url}
\usepackage{pifont}
\usepackage{paralist}
\usepackage{enumitem}
\usepackage{balance}
\usepackage{comment}
\usepackage{multirow}
\usepackage{upgreek}
\usepackage[ruled,vlined]{algorithm2e}
\usepackage[fleqn]{nccmath}
\usepackage{epsfig,amsfonts,amsmath,amssymb,graphics, color,wrapfig}
\usepackage{epstopdf}
\usepackage{epsfig}

\usepackage[normalem]{ulem}

\usepackage{epsfig, subfigure, amsmath, amssymb}

\usepackage{epstopdf}
\usepackage{balance}
\usepackage{color}
\usepackage{url}
\usepackage{pifont}
\usepackage{paralist}
\usepackage{commath}
\usepackage{enumitem}
\usepackage{balance}
\usepackage{comment}
\usepackage{xcolor}
\usepackage{soul}
\usepackage{graphicx}
\usepackage{mwe}
\usepackage{caption}
\usepackage{hyperref}

\usepackage{pifont}
\usepackage{graphicx}
\usepackage{xcolor}
\usepackage{enumitem}
\usepackage{multicol}

\definecolor{red}{rgb}{1,0,0}
\definecolor{green}{rgb}{0,1,0}
\definecolor{blue}{rgb}{0,0,1}
\definecolor{aqua}{rgb}{0.7,0,0.7}
\definecolor{orange}{rgb}{1,0.64,0.06}

\def\BibTeX{{\rm B\kern-.05em{\sc i\kern-.025em b}\kern-.08em
    T\kern-.1667em\lower.7ex\hbox{E}\kern-.125emX}}

\begin{document}

\title{A Novel Framework for Threat Analysis of Machine Learning-based Smart Healthcare Systems}
\author{
\IEEEauthorblockN{Nur Imtiazul Haque\IEEEauthorrefmark{1}, Mohammad Ashiqur Rahman\IEEEauthorrefmark{1}, Md Hasan Shahriar\IEEEauthorrefmark{1} \\ Alvi Ataur Khalil\IEEEauthorrefmark{1} and Selcuk Uluagac\IEEEauthorrefmark{2}}
\IEEEauthorblockA{\IEEEauthorrefmark{1}Analytics for Cyber Defense (ACyD) Lab, \IEEEauthorrefmark{2}Cyber-Physical Systems Security Lab\\
Department of Electrical and Computer Engineering\\
Florida International University, Miami, USA\\
	\{ nhaqu004, marahman, mshah068, akhal042, suluagac\}@fiu.edu
	}
}

\maketitle

\begin{abstract}
Smart healthcare systems (SHSs) are providing fast and efficient disease treatment leveraging wireless body sensor networks (WBSNs) and implantable medical devices (IMDs)-based internet of medical things (IoMT). In addition, IoMT-based SHSs are enabling automated medication, allowing communication among myriad healthcare sensor devices. However, adversaries can launch various attacks on the communication network and the hardware/firmware to introduce false data or cause data unavailability to the automatic medication system endangering the patient's life. In this paper, we propose SHChecker, a novel threat analysis framework that integrates machine learning and formal analysis capabilities to identify potential attacks and corresponding effects on an IoMT-based SHS. Our framework can provide us with all potential attack vectors, each representing a set of sensor measurements to be altered, for an SHS given a specific set of attack attributes, allowing us to realize the system's resiliency, thus the insight to enhance the robustness of the model. We implement SHChecker on a synthetic and a real dataset, which affirms that our framework can reveal potential attack vectors in an IoMT system. This is a novel effort to formally analyze supervised and unsupervised machine learning models for black-box SHS threat analysis.
\end{abstract}


\section{Introduction}
\label{sec:introduction}
Traditional healthcare systems rely heavily on hospitalization, specialized consultation, and nursing that require a lot of human interventions. It introduces delayed or incorrect treatment resulting in increased treatment cost or human mortality. In the U.S., the aggregated patient treatment cost was almost \$3.81 trillion in 2019~\cite{healthaffairs2020}, which is projected to be much higher in recent days, especially due to COVID-19~\cite{healthaffairs2020-COVID19}. An automated healthcare system can certainly reduce this cost and the death rate. Contemporary internet of medical things (IoMT) technology has brought a radical revolution in the healthcare domain by enhancing the reliability of remote patient monitoring, increasing efficiency of the medical sensors, and eliminating latency between disease detection and medication~\cite{joyia2017internet}. A growing number of researches have been conducted toward using IoMT technology for ubiquitous data acquisition, timely processing, and wireless data distribution in the healthcare field~\cite{wang2010wireless,agarwal2011input}. 

The smart healthcare system (SHS) is a modern cyber-physical system (CPS) that continuously collects data from the IoMT sensor network connected to the human body, processes them accordingly for making required control decisions, and triggers implantable medical devices (IMDs) for real-time medication and treatment. Currently, healthcare facilities are more efficient, accessible, and personalized as the SHS is ameliorating disease diagnostic tools, treatment for patients, and healthcare devices, thus improving the quality of lives~\cite{demirkan2013smart}. However, an SHS requires processing a lot of historical data to identify anomalous sensor measurements. The data related to healthcare and medication are affluent. They can be utilized to reveal intricate patterns of dependency between various vital signs of the human body for accurate and precise disease classification. For faster processing, an SHS integrates the concept of IoMT with big data, cloud computing, and artificial intelligence~\cite{tian2019smart}.

With the advent of SHSs, smart medical devices are being exposed to numerous attack points, susceptible to potential threats. Cyberattacks on the healthcare industry are rapidly growing~\cite{threatbrief2017}. SHS devices/IMDs often have vulnerabilities that expose significant threats~\cite{cbsnews2018}. According to a recent report, 53\% of the healthcare organizations were attacked between October 2018 and October 2019~\cite{ponem2019}. Most popular cyberattacks in the healthcare systems includes hardware Trojan~\cite{wehbe2017novel}, malware (e.g., Medjack~\cite{Storm2015medjack}), Sybil attacks using either hijacked IoMT \cite{almogren2020ftm} or single malicious node \cite{bapuji2018internet}, DoS attacks \cite{deshmukh2015understanding}, 
and man-in-the-middle (MITM) attacks~\cite{pournaghshband2012securing}.  At least 20\% of the medical device manufacturers experienced ransomware or malware attacks in the last 20 months \cite{naveen2020mal}. 
So, it is imperative to study the vulnerabilities of an SHS before deploying it. 

\begin{figure*}[t]
\centering
\includegraphics[width = 0.95\textwidth]{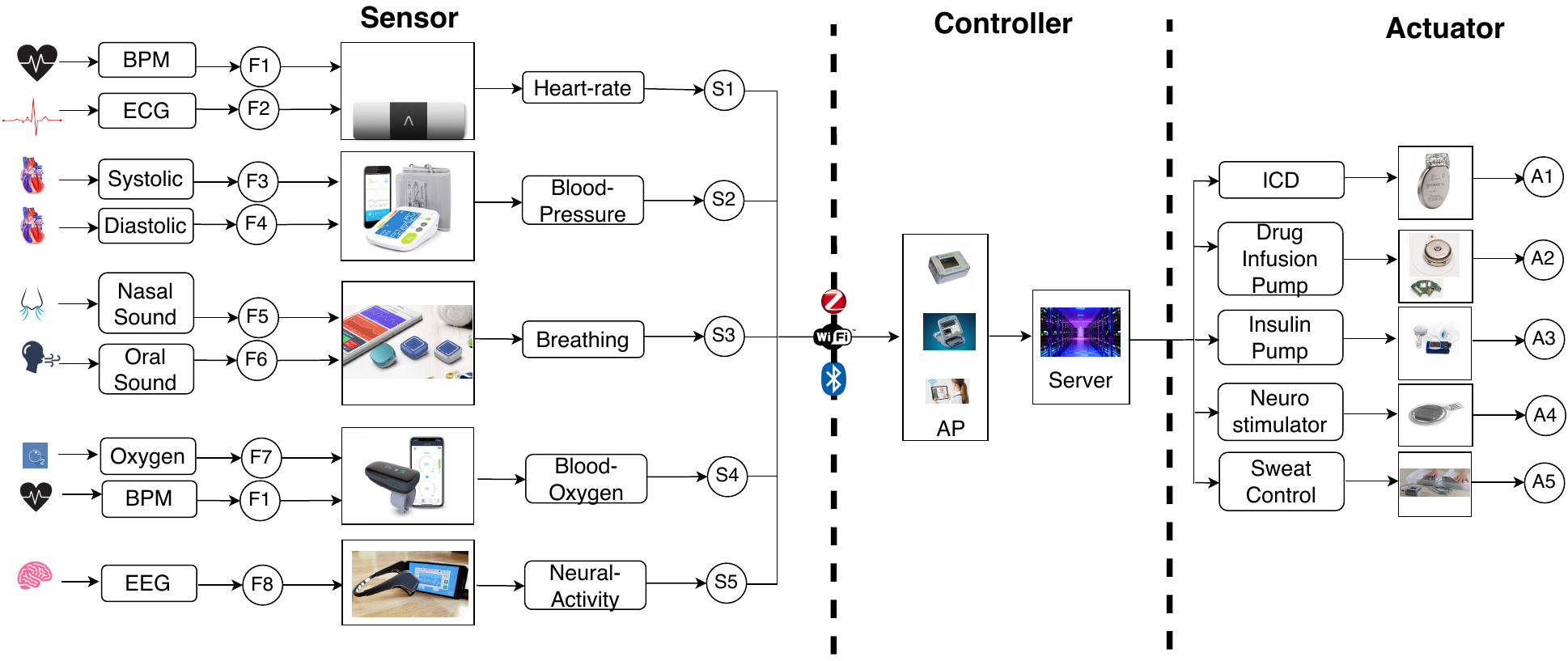}
\centering
\caption{Sensor behavioral model in SHS}
\label{fig:sensor_behavioral_model}
\end{figure*}
In this paper, we propose an automated threat analysis framework, named SHChecker, for SHSs that uses a machine learning (ML)-based disease classification model (DCM) to deliver real-time treatment. SHChecker analyzes the underlying decision-making model of SHS by investigating the possible attacks that can be deployed by minimal alteration of sensor values. For safety-critical CPSs like SHSs, failures can raise the possibility of life-threatening events. That's why, to increase reliability, SHSs often employ data validation or anomaly detection systems. Hence, we consider a clustering-based anomaly detection model (ADM) in the SHS due to its real-time detection capability~\cite{ariyaluran2019clustering}. The ADM can learn the pattern of sensor measurement relationships by analyzing a massive quantity of data. The proposed SHChecker framework can assess potential attack vectors of an SHS that uses ML algorithms, such as decision tree (DT), logistic regression (LR), or artificial neural network (ANN) for classifying diseases and K-means or density-based spatial clustering of applications with noise (DBSCAN) clustering algorithms for detecting anomalies. Our work mainly focuses on quantifying the associated threat that can be performed by minor alteration in one or more sensor measurements of an SHS. We verify our model's efficacy using the University of Queensland Vital Signs (UQVS) dataset and our synthetic dataset using several performance metrics~\cite{liu2012university}.

In summary, our contributions are as follows:
\begin{itemize}
    \item Considering various ML algorithms, we model a real-time SHS by deploying a DCM and a corresponding ADM.
    \item We formally represent the attack model for an SHS using a set of flexible attack attributes that specifies an attacker's capabilities and the attack target.
    \item We build SHChecker, a threat analysis framework that can identify potential attack vectors for an ML model-based SHS. The source code can be found in github~\cite{shchecker2020}.
    \item We conduct experiments with SHChecker using real and synthetic data to analyze the attack characteristics and evaluate the tool's scalability. 
\end{itemize}

The rest of the paper is organized as follows: we provide an overview of the SHS and its vulnerabilities along with other necessary background information in Section~\ref{sec:background}. We present the SHChecker's architecture in Section~\ref{sec:proposed_framework}. In Section~\ref{sec:technical_details}, we discuss the technical details of the framework, which includes the machine learning-based models for constraint generation and a formal model for threat analysis considering adversary's goals, capabilities, and methodology. We also provide example case studies in section~\ref{case_study}. Moreover, we discuss our SHS testbed implementation in Section~\ref{sec:implementation}. Then, we evaluate our proposed framework by running experiments on a synthetic dataset and a real dataset and present the results in Section~\ref{sec:evaluation}. Finally, we conclude the paper in Section~\ref{Sec:conclusion}.
\label{sec:introduction}

\section{Background}
\label{sec:background}
This section provides an overview of the SHSs and some other related information to motivate the work and facilitate the reader's comprehension.
\subsection{IoMT-based SHS}
IoMT based SHS is a game-changer for the medical field concerning consultation accuracy and cost reduction related to human labor. The enormous amount of medical data enables researchers to perform statistical analyses of diseases and medication patterns. So, with the introduction of IoMT in the healthcare field, more attention has been paid to developing ubiquitous data accessing solutions to acquire and process data from decentralized data sources~\cite{kyusakov2012integration, wang2010service, pereira2007widom}. In the IoMT network, data are sent to a remote server to analyze and take control decisions due to the lack of processing capability of medical sensors and implantable medical devices (IMDs).

This paper considers an IoMT applied SHS incorporating a wireless body sensor network, ML-based control system, and IMD-based actuators.Fig.~\ref{fig:sensor_behavioral_model} demonstrates an SHS that can deliver medicine in real-time with a closed-loop decision control system without requiring any human involvement. In an SHS, patients are continuously monitored by the sensors attached to their bodies. These sensors deliver their observed measurements to the controller using some wireless communication protocols, e.g., WiFi, Bluetooth, Zigbee, etc. 

The controller takes thee measured values, makes decisions based on them, and sends the control commands to the IMDs to deliver the necessary treatment to the patient. For example, Dario's blood glucose monitoring system continuously advertises blood glucose value to the controller (Table~\ref{table:devices}).  The controller checks whether the vital signs of the patient are within normal ranges~\cite{hoskins2016}. If the controller 
figures out that the patient needs emergency insulin delivery, it notifies the responsible insulin pump implanted inside the patient's body to inject the proper amount of medication. 
\begin{table*}[t!]
\caption{Devices and parameters considered for monitoring health condition. }
\label{table:devices}
\centering
\small
\resizebox{1\textwidth}{!}{
\begin{tabular}{|l|l|l|l|l|}
\hline
\multicolumn{1}{|c|}{\textbf{Vital Signs}} & \multicolumn{1}{c|}{\textbf{Model}}       & \multicolumn{1}{c|}{\textbf{Feature Parameter Value}}                                & \multicolumn{1}{c|}{\textbf{Database}}            & \multicolumn{1}{c|}{\textbf{Ref.}} \\ \hline
Heart Rate                                            & KardiaMobile 6L                                 & 60-100 beats per minute                                                               & Fetal ECG Synthetic Database, Data.Gov                      & \cite{diabetes}                                  \\ \hline
Blood Pressure                                            & Greater Goods                                 & Systolic (120 mm Hg) and Diastolic (80 mm Hg)                                                                & Fetal ECG Synthetic Database, Data.Gov                      & \cite{diabetes}                                  \\ \hline

Blood Glucose                                         & Dario        & 70-130 mg/dl                                                                         & UCI ML database of diabetes         & \cite{martin2012higher}                                  \\ \hline
Blood Oxygen                                          & iHealth Air Wireless Pulse Oximeter       & Oxygen Saturation level $\ge$ 94\%                                                   & Pattern Analysis of Oxygen Saturation Variability & \cite{bhogal2017pattern}                                  \\ \hline
Respiratory Rate                                      & QuardioCore                               & 12-20 Breaths per minute                                                             & BIDMC PPG and Respiration Dataset                 & \cite{pimentel2017toward}                                  \\ \hline
Blood Alcohol                                         & Scram Continuous Alcohol Monitoring (Cam) & 0.08 g/dl                                                                            & StatCrunch dataset                                & \cite{fell2014effectiveness}                                  \\ \hline
\end{tabular}}
\end{table*}

\subsection{Machine Learning in SHS}
ML algorithms are being used in critical applications where they drive decisions with enormous personal, organizational, or societal impact, and healthcare is one of them ~\cite{brameier2001comparison}.  In a growing industry of healthcare sensors that continuously gather a plethora of health data, the prevalence of using machine learning to analyze these data is gaining momentum. Our considered SHS model uses a DCM and an ADM. The DCM uses a supervised ML algorithm that can label patient data accurately in real-time. ANN-based deep learning model and rules-based models like DT demonstrate moderate performance in disease classification~\cite{kaur2006empirical, srinivas2010applications}. According to this classification, the controller makes decisions. The ADM uses an unsupervised ML technique to learn the complex relationship between various features and detect anomalous measurements. Usually, anomaly detection mechanisms take longer to execute, as most of the time, they do not provide any explicit model. The ADM verifies DCM-provided decisions.

\subsection{Cyber Attacks in SHSs}
Malware and man-in-the-middle (MITM) attacks are predominant in SHSs. One of the most recent malware attacks named ``MEDJACK'' (Medical Device Hijacking) exploits healthcare systems by placing malware within the IoMT networks~\cite{Storm2015medjack}. MEDJACK is a stealthy cyber-attack that utilizes the concept of polymorphic malware by constantly escalating its capability, making it very difficult to get revealed. By creating a backdoor behind the firewall, MEDJACK gains access to the network without being detected. 

MITM is a cyber-attack where an adversary illegally gets into the communication between two authorized parties and eavesdrops on the transmitted data or corrupts it.
Bluetooth-enabled medical devices exhibit potential vulnerabilities in sensor networks. Pournaghshband et al.~\cite{pournaghshband2012securing} demonstrate the feasibility of launching an MITM attack in a Bluetooth-enabled pulse oximeter, which confirms that medical sensors can be compromised. They reverse-engineered a Nonin Onyx II 9550 fingertip pulse oximeter, which can gauge blood oxygen saturation level and pulse rate. The MITM attacks on wireless links can be performed in various ways, e.g., by jamming Bluetooth pairing with devices or access points (APs).

\subsection{Threat Model}
Threat modeling is the process of potential threat recognition and security measures to protect a valuable system. We aim to develop a threat analysis model considering data corruption, MITM, and malware injection attack that can compromise SHS sensor values without getting detected by the system to prevent providing intended service from a system or deliver the wrong service. In our model, we consider a powerful attacker who can eavesdrop and alter sensitive safety-critical data in such a way that provokes the control system to make a wrong decision.

An attacker's capabilities specify the appropriate ways (e.g., knowledge, time, expertise, and tools) and opportunities (e.g., enough time to perform the attack) to exploit the vulnerabilities that can materialize possible threat~\cite{ben2013using}. Our threat model identifies potential attack vectors, injecting which can significantly affect the system's consistency and increase the probability of inaccurate medication delivery.
\label{sec:background}

\section{SHChecker Framework}
\label{sec:proposed_framework}

\begin{figure}[t] 
\centering
\includegraphics[width=0.9\columnwidth]{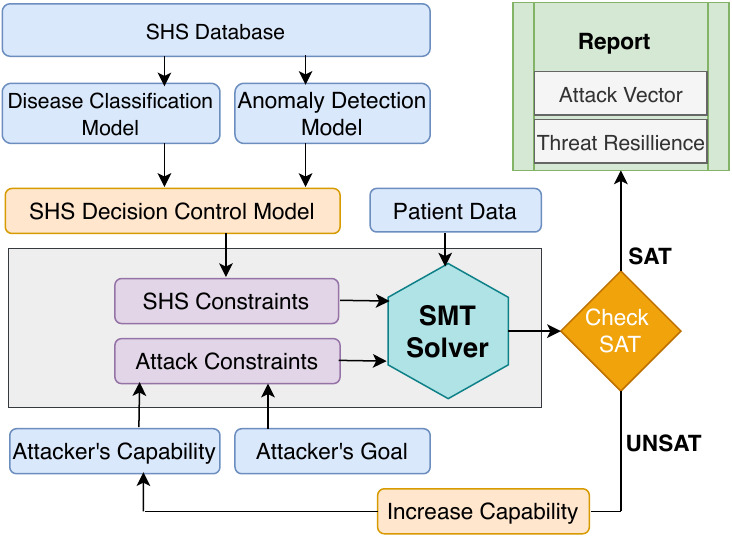}
\caption{The architecture of SHChecker.}
\label{fig:framework}
\end{figure}

The underlying architecture of our proposed SHS threat analysis framework, SHChecker is presented in Fig.~\ref{fig:framework}. In this framework, we first feed the smart healthcare dataset to train two machine learning models. The DCM is trained to label the patients, and the ADM verifies the consistency of the sensor data for that specific level. The outputs from these two models are used in the SHS decision control model. The decision control model processes the inputs from the machine learning models into decision boundaries and generates necessary constraints accordingly. Then, an SMT solver takes all the constraints associated with the SHS model and the attacker's capability and goal, along with the sensor/measurement data of a patient as input, and encode the analysis of the attack feasibility as a constraint satisfaction problem (CSP). 

The satisfiability modulo theory (SMT) solver utilizes various background theories to solve this CSP. When it returns a SAT result, it implies that the given set of constraints is satisfied for the patient data into consideration. A SAT result reports an attack vector that includes a set of values to be injected to a set of sensor measurements for misclassifying the given patient data based on the attacker's goal. The framework chooses the attacker's capability in a systematic way so that we can receive a SAT result at the minimum capability. In this case, the attack vector also implies that the attacker cannot succeed in his goal if his capability is lower than the framework reported capability. Hence, the system can be defined as {\slshape threat resilient} till that attack's capability in terms of cybersecurity.

An UNSAT result by the SMT solver signifies that the attack is not possible for attaining the attacker's goal based on the given capabilities. In such a case, the framework increases the attacker's capability by providing him/her with access to more resources, reexamines the attack feasibility, and repeats the process until being successful in finding an attack vector or attacker's capability can no longer be increased.
\label{sec:proposed_framework}

In this section, we present the technical details of the SHS threat analysis framework. The framework consists of two major functional components: machine learning models and a formal analysis model.
\subsection{Machine Learning Model} \label{Sec:ml_model_det}
Our comprehensive framework performs a threat analysis of two different types of ML-based models: DCM and ADM. The DCM considers three widely used algorithms, namely decision tree, logistic regression, and neural network. For ADMs, DBSCAN and K-means clustering-based models are adopted.

\noindent\textbf{Decision Tree (DT) Algorithm:}
DT is a popular classification model for its comprehensive, fast, easy to use and versatile nature. DT produces a tree data structure like decision rules to devise a category based on the dataset features. Determining the feature selection order and finding the best splitting points are the main challenges to generate the inference rules. These challenges are referred to as attribute selection in terms of machine learning. Various metrics are used for attribute selection, such as information gain, Gini index, etc~\cite{quinlan1986induction}. In our case, we implement our decision tree using the CART (Classification and Regression Trees)-based algorithm and consider the Gini index metric for attribute selection.

\noindent\textbf{Logistic Regression (LR) Algorithm:}
LR model is an efficient baseline model for solving classification problem, which works better for drawing accurate decision boundary between various classes when the relationship between features and labels are straightforward. As LR is mainly designed for solving binary classification problems, selecting a multiclass classification approach arises a significant amount of challenges. Moreover, LR uses various solvers to optimize cost function for finding the best decision boundary. In our framework, we utilize the one versus rest scheme for performing multiclass classification and limited-memory Broyden–Fletcher–Goldfarb–Shanno solver for optimizing our classification's cost function problem~\cite{saputro2017limited}. 

\noindent\textbf{Artificial Neural Network (NN) Algorithm:}
Artificial Neural network algorithms can replicate the human brain and perform multiple tasks parallel to retaining system performance. Most importantly, unlike LR, NN can discover intricate patterns from data. NN model works remarkably well for multiclass classification problems, although training such a model requires a lot of tuning such as learning rate, batch size, number of hidden layer, etc~\cite{beale1996neural}. SHChecker framework uses a feed-forward NN that optimizes its cost function using an adam optimizer.

\noindent\textbf{DBSCAN Algorithm:}
For the ADM, we consider the DBSCAN algorithm as it shows promising performance for finding anomalies \cite{ccelik2011anomaly}. DBSCAN splits the dataset into several clusters using two hyperparameters, epsilon and minpoints (the minimum number of points to create a cluster). Any point that does not fit into any cluster is denoted as a noisy point. Our framework groups the good data points based on DT labels and finds the optimal values for the parameters to cluster all the good data points. However, DBSCAN does not give any explicit decision boundary for those clusters.

\noindent\textbf{K-means Algorithm:}
K-means algorithm is another popular unsupervised clustering algorithm that creates a given number of the cluster over the data samples applying the data's mean. Unlike the DBScan algorithm, determining the optimal number of clusters, $k$ is a big challenge in the K-means algorithm. However, the K-means algorithm does not come up with an explicit decision boundary as well. 
For both ADMs, we use the Euclidean distance metric.

All of these algorithms leverage the relationship between the features. The decision rules from the DCM are used to produce constraints associated with the SHS. However, to generate data validity of cluster-derived constraints from the anomaly detection algorithm, we need to define the decision boundary formally. 
A concave hull algorithm can be used to get a tight bound for the clusters~\cite{moreira2007concave}. The concave hull algorithm uses a k-nearest neighbor-based approach to fit the data points in a best-described polygon concave polygon that can be smoothed by a hyperparameter, k~\cite{guo2003knn}.


\begin{table*}[]
\scriptsize
\caption{Modeling Notations\label{tab:modeling_parameters}}
\begin{tabular}{|p{1.8cm}|p{2.2cm}|p{1.5cm}|p{9.2cm}|p{1cm}|}
\hline
\textbf{Model Type}                                                                         & \textbf{ML Model}                                                               & \textbf{Symbol}           & \textbf{Description}                                                                               & \textbf{Type} \\ \hline
\multirow{6}{*}{All}                                                                        & \multirow{6}{*}{All}                                                            & $n_s$                     & Number of sensors                                                                                  & Integer       \\ \cline{3-5} 
                                                                                            &                                                                                 & $n_l$                     & Number of labels                                                                                   & Integer       \\ \cline{3-5} 
                                                                                            &                                                                                 & $a,b$                     & Sensors                                                                                            & String        \\ \cline{3-5} 
                                                                                            &                                                                                 & $j$                       & Patient's status                                                                                   & String        \\ \cline{3-5} 
                                                                                            &                                                                                 & $\mathcal{S}$             & Set of all sensors                                                                                 & Set           \\ \cline{3-5} 
                                                                                                &                                                                                 & $\mathcal{L}$             & Set of all possible statuses for a patient                                                         & Set           \\ \hline
\multirow{9}{*}{\begin{tabular}[c]{@{}l@{}}Disease \\ Classification \\ Model\end{tabular}} & \multirow{2}{*}{Decision Tree}                                                  & $Np^f$                    & Set of all nodes in path f                                                                         & Set           \\ \cline{3-5} 
                                                               &                                                                                 & $Np$             &  Set of all nodes                                                                                 & Set\\ \cline{3-5}                                             &                                                                                 & $\mathrm{Pth}$                      & Set of all paths from root to leaf                                                                   & Set           \\ \cline{2-5} 
                                                                                            & \multirow{5}{*}{\begin{tabular}[c]{@{}l@{}}Logistic \\ Regression\end{tabular}} & $\theta_{gi}$             & Model parameter (weight) for i-th sensor and g-th patient status                                   & Real          \\ \cline{3-5} 
                                                                                            &                                                                                 & $\theta_{g}$              & Set of model parameter (weight) for g-th patient status                                            & Set           \\ \cline{3-5} 
                                                                                            &                                                                                 & $\theta$                  & Set of all model parameter (weight)                                                                & Set           \\ \cline{3-5} 
                                                                                            &                                                                                 & $intercept_g$             & Model parameter (bias) for g-th patient status                                                     & Real          \\ \cline{3-5} 
                                                                                            &                                                                                 & $intercept$               & Set of all model parameter (bias)                                                                  & Set           \\ \cline{2-5} 
                                                                                            & \multirow{2}{*}{Neural Network}
                                                                                            & $NL_i$                    & Set of Nodes in i-th layer                                                                  & Set           \\ \cline{3-5} 
                                                                                            
                                   &                                                                                 & $NL$                      & Sets of all layers                                                                  & Sets           \\                                                                      \hline
\multirow{6}{*}{\begin{tabular}[c]{@{}l@{}}Anomaly \\ Detection \\ Model\end{tabular}}      & \multirow{6}{*}{DBScan, K-means}                                                & $\mathcal{C}_k^{a, b,j}$  & $k$-th cluster for the sensor pair $(a,b)$ for a patient with status, $j$                          & Cluster       \\ \cline{3-5} 
                                                                                            &                                                                                 & $\mathcal{C}^{a, b,j}$    & Set of all clusters  for the sensor pair $(a,b)$ for a patient with status, $ j$                   & Set           \\ \cline{3-5} 
                                                                                            &                                                                                 & $\mathit{Ls}_l^{a, b, j}$ & $l$-th line segment of the clusters for the sensor pair $(a,b)$ and the patient's status, $j$      & Tuple         \\ \cline{3-5} 
                                                                                            &                                                                                 & $\mathit{Ls}^{a, b, j}$   & Set of all line segments of the clusters for the sensor pair $(a,b)$ and the patient's status, $j$ & Set           \\ \cline{3-5} 
                                                                                            &                                                                                 & $\mathcal{C}_{all} $      & Sets of all clusters                                                                               & Sets          \\ \cline{3-5} 
                                                                                            &                                                                                 & $ \mathrm{Ls}_{all}$      & Sets of all line segments                                                                          & Sets          \\ \hline
\end{tabular}
\end{table*}

\subsection{Formal Analysis Model} \label{Sec:technical_detail}
Our proposed framework performs a formal analysis by modeling the problem as a CSP that considers two kinds of constraints associated with modeling (i) the SHS and (ii) attacks. Table~\ref{tab:modeling_parameters} describes the notation using in this paper for formal modeling.

\vspace{12pt}
\noindent{\slshape \bf Formal Modeling of SHS:}

Let us assume that $\mathcal{P}$ is the set of sensor measurements of a patient with a patient status/label, $j$. To verify the authenticity of the sensor measurements and the label, we use the SHS model. If the data satisfies both disease classification and ADM-driven constraints, we consider it as validated for the patient's status.

\vspace{3pt}
\noindent\textbf{Derivation of Classifier-Driven Constraints:}

\noindent\textbf{DT Constraints:}
A decision tree model returns an inference hierarchical rules-based model, from which formal model constraints acquisition to represent the model is quite straightforward. A boolean function $\mathit{inference}(\mathcal{P},{j})$ returns $True$ if sensor values are consistent with label $j$ for DT inference rules.

\begin{figure}[!t]
\centering
\includegraphics[width=0.6\columnwidth]{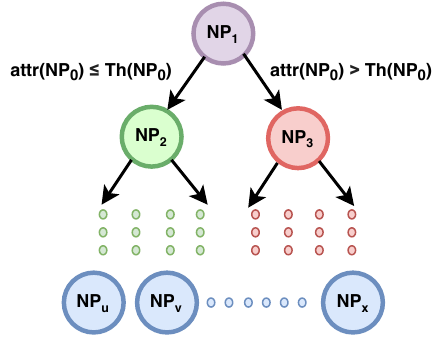}
\caption{Decision Tree}
\label{fig:dt}
\end{figure}

DT contains several nodes starting from the root node. Each node of the tree consists of an attribute which denotes a sensor measurement using which the tree is split at that point with a threshold value. A patient's sensor measurement of that particular attribute having greater than the threshold follows the right path at that particular node, otherwise it follows the left path. This attribute and the threshold value generates a rule as shown in the Equation~\ref{eq:dt_rule}. Fig.~\ref{fig:dt} demonstrates a decision tree model for understanding the DT constraints.

\begin{equation} \label{eq:dt_rule}
\mathit{rule}( \mathcal{P}_{a},d,e) = 
\begin{cases}
  \mathcal{P}_{a} \leq Th(d) & \text{if e is a left node of d} \\
  \mathcal{P}_{a} > Th(d) & \text{if e is a right node of d}
\end{cases}
\end{equation}
Here, $e$ is a immediate child of $d$ and $a = \mathit{attr}(d)$

The whole tree is divided into multiple paths, from the root to the leaf. Each path has a label and set of rules along its way. Equation~\ref{eq:dt_rules} demonstrates the process of determining rules from a set of rules along a path.
%
\begin{equation} \label{eq:dt_rules}
  \mathit{rules}(\mathcal{P},f) = \bigwedge\limits_{i=1}^{|Np^f|-1}  \mathit{rule}(\mathcal{P}_{attr(Np^f_i)},Np^f_i,Np^f_{i+1})
\end{equation}

The decision tree assigns a label $j$ as patient status if and only if sensor measurements associated with that patient satisfy all inference rules along a path having label $j$. Here, $j$ is the label of the last node of the path.
%
\begin{equation}
\begin{split}
    \mathit{inference}(\mathcal{P},j) \iff \exists_{f \in \mathrm{Pth}} label(f) = j \land \mathit{rules}(\mathcal{P},f)
\end{split}
\end{equation}

\noindent\textbf{LR Constraints:}
A logistic regression model assigns some probability values to each patient status for a patient. The label that gets the highest probability after applying the softmax function is selected as that patient status. The model parameters, in this case, have been obtained by minimizing a cost function for optimal decision boundary using maximum log-likelihood. The inference constraints for LR is shown in Equation~\ref{eqn:lr_constraint}.
%
\begin{equation} \label{eqn:lr_constraint}
\begin{split} 
    &\mathit{inference}(\mathcal{P},j)\iff \\ & ~~ \underset{g}{\arg\max} \frac{ exp( (\sum_{i=1}^{n_s} \mathcal{P}_g\theta_{gi})+ \epsilon_g ) }
    { \sum_{h=1}^{n_l} exp ((\sum_{i=1}^{n_s} \mathcal{P}_h\theta_{hi}) + \epsilon_h ) } = j
\end{split}
\end{equation}


\noindent\textbf{NN Constraints:}
A neural network model comprises a number of layers categorized by an input layer, one or more hidden layers, and one output layer. The input of each node at any layer except the input layer is calculated from the previous layer's output, weights, and bias.

\begin{figure}[!b]
\centering
\includegraphics[width=0.66\columnwidth]{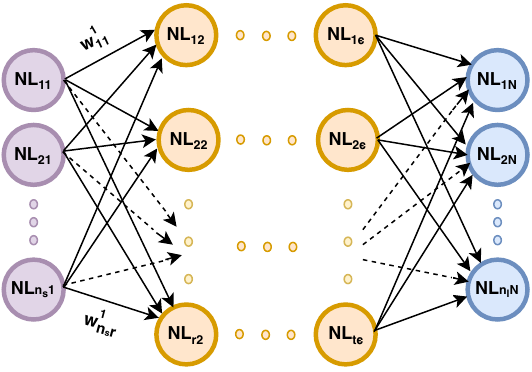}
\caption{Neural Network}
\label{fig:nn}
\end{figure}

Suppose, $N$= $|\mathrm{NL}|$, number of layers in the model.
%
\begin{equation}
\begin{split}
    \forall_{m \in (1,N]} input(\mathrm{NL}_{mn}) = \sum_{o=1}^{|\mathrm{NL}_{m-1}|}(output(\mathrm{NL}_{(m-1)o}) \times \\  W_{on}^{m-1}) + b_{mn}
\end{split}    
\end{equation}

The input and output of layer 1 are the sensor measurement values as shown in the Equation~\ref{eq:layer_1}. Fig.~\ref{fig:nn} demonstrates a neural network model of $N$ layers where the last hidden layer is denoted by $\epsilon$ i.e. $\epsilon = N-1$.
%
\begin{equation} \label{eq:layer_1}
\begin{split}
    input(\mathrm{NL}_{1}) = output(\mathrm{NL}_{1}) =  \mathcal{P} 
\end{split}    
\end{equation}

For calculating each node's output, input values of a particular node are passed through some complex activation function like relu, tanh etc.
%
\begin{equation}
\begin{split}
    \forall_{m \in (1,N]}output(NL_{mn}) = activation(input(NL_{mn}))    
\end{split}
\end{equation}

Label j is assigned to the patient in consideration if and only if softmax function outcome of j-th output node gets higher value than the other output nodes.
%
\begin{equation}
\begin{split}
    \mathit{inference}(\mathcal{P},j) \iff \underset{g}{\arg\max} \frac{ exp(input(NL_{ng}) }
    { \sum_{q=1}^{n_l} exp ( input(NL_{nq})  } = j
\end{split}
\end{equation}

\vspace{3pt}
\noindent\textbf{Derivation of ADM-Driven Constraints:}

\noindent\textbf{DBSCAN Algorithm-Based constraints:}
To validate the consistency of a set of measurements, we use the DBSCAN algorithm-driven constraints. 

We consider consistency between the combination of all pair of sensor measurements instead of all sensor measurements together to face the challenge of obtaining constraints in high dimensional space. Because most of the clusters do not satisfy the requirement of constraint acquisition in high dimensional space due to lack of sufficient cluster data points, we find these constraints according to the logical functions of checking if the measurements are within that specific label's clusters. To facilitate the reader, we explain this concept below with a simple example.

\begin{figure}[!t]
    \centering
    \includegraphics[width=0.85\columnwidth]{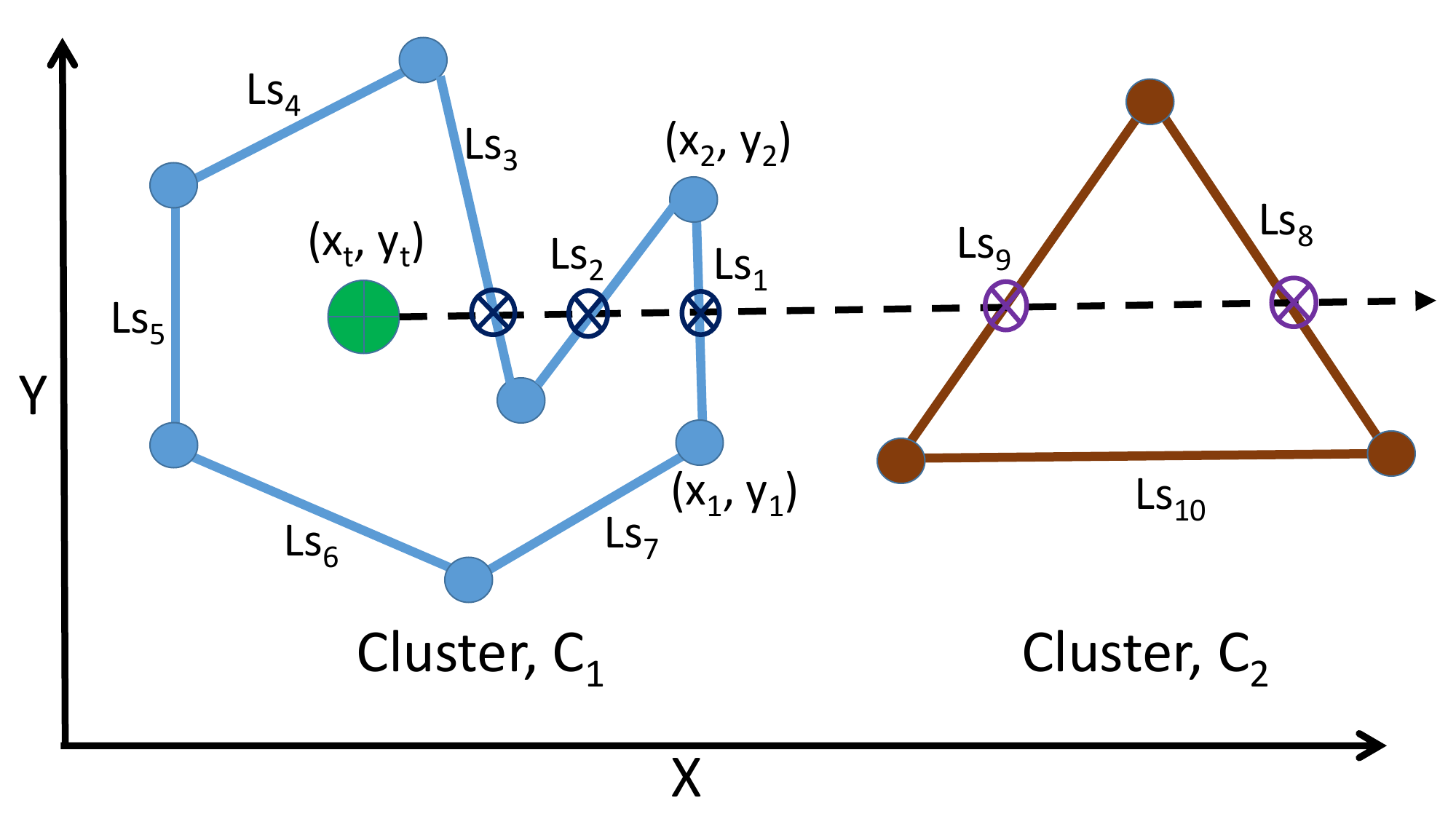}
    \caption{Logic behind checking if a point is inside a polygon cluster in DBSCAN Algorithm.}
    \label{fig:toy_example}
\end{figure}

Fig.~\ref{fig:toy_example} shows two clusters ($C_1$ and $C_2$) in a 2D data plane where $C_1$ consists of seven line segments ($\mathit{Ls}_1, \mathit{Ls}_2, \ldots, \mathit{Ls}_7$) and $C_2$ consists of three line segments ($\mathit{Ls}_8, \mathit{Ls}_9$, and $\mathit{Ls}_{10}$).  
We denote the end points of any line segment ($\mathit{Ls}_i$) are $(x_a^i, y_a^i)$ and $(x_b^i, y_b^i)$, where $y_b^i \geq y_a^i$. To validate the consistency of a data point $(x, y)$, we use the following logical functions:

\begin{itemize}[leftmargin=*]
\vspace{6pt}
\item $\mathit{inRangeOfLineSegment}$ $(x, y, \mathit{Ls}_i)$: This function checks whether the point is within the vertical range of the line segment, $\mathit{Ls}_i$. 
\begin{equation}
\label{inRangeOfLineSegment_0}
\mathit{inRangeOfLineSegment}(x, y, \mathit{Ls}_i) \iff y_a^i < y \leq y_b^i
\end{equation}

Thus, according to Fig.~\ref{fig:toy_example}, we can say that,\\ $\mathit{inRangeOfLineSegment}  (x_t, y_t, \mathit{Ls}_1)$ returns $True$ for a point ($x_t,y_y$) as $y_t$ is within the range of $y_1^1$ and $y_2^1$. However, for line segment $Ls_4$, $\mathit{inRangeOfLineSegment}(x_t, y_t, Ls_4)$  return $False$ as $y_t$ is not in the vertical range. 

\vspace{6pt}
\item $\mathit{leftOfLineSegment}(x, y, \mathit{Ls}_i)$: This function checks if the point $(x, y)$ is on the left side of the line segment, $\mathit{Ls}_i$. 
\begin{equation}
\label{left_of_line_segment_0}
\begin{split}
&\mathit{leftOfLineSegment}(x,y, \mathit{Ls}_i) \iff \\
& ~~~~(x(y_a^i - y_b^i) - y(x_a^i - x_b^i) - (x_a^i y_b^i - x_b^i y_a^i)) < 0
\end{split}
\end{equation}

For cluster $C_1$ in Fig.~\ref{fig:toy_example},  $\mathit{leftOfLineSegment}(x_t,y_t, \mathit{Ls}_1)$ returns $True$ while $\mathit{leftOfLineSegment}(x_t,y_t, \mathit{Ls}_5)$ returns $False$.

\vspace{6pt}
\item $\mathit{intersect}(x, y, Ls_i)$: An imaginary line is drawn from the point $(x, y)$ to the right side, which is also parallel to the $x$ axis. The function $\mathit{intersect}$ determines if the imaginary line intersects the line segment, $\mathit{Ls}_i$. The imaginary line intersects the line segment only when the point $(x,y)$ is within the range of the line segment and also located on the left side of it. The function is formalized as follows:
\begin{equation}
\label{split_0}
\begin{split} 
\mathit{intersect} & (x,y, \mathit{Ls}_i) \iff \mathit{inRangeOfLineSegment}(x,y, \mathit{Ls}_i) \land \\ 
&\mathit{leftOfLineSegment}(x, y, \mathit{Ls}_i)
\end{split}
\end{equation}
From Fig.~\ref{fig:toy_example}, it is obvious that $\mathit{intersect}(x_t, y_t, Ls_i)$ returns $True$ only for line segments $1, 2, 3, 8,$ and $9$.

\vspace{6pt}
\item $\mathit{withinCluster}(x, y, C_k)$: This function returns $True$ if the data point $(x,y)$ is within the cluster, $C_k$. For all the (boundary) line segments ($\mathit{Ls}_i$) of $C_k$, the function calculates $\mathit{intersect}(x, y, \mathit{Ls}_i)$ and performs the $XOR$ operation on them. If there are an odd number of intersections with the line segments, $\mathit{withinCluster}$ returns $True$ as the $XOR$ operation on one or more $False$ values and an odd number of $True$ values results  $True$. Thus, we define the function as:
\begin{equation}
    \begin{split}
     &\mathit{withinCluster}(x, y, C_k) \iff \\
     & ~~\bigoplus_{1 \leq i  \leq \mathit{numOfLs}_{C_k}} (\mathit{intersect}(x,y, \mathit{Ls}_i) \land \mathit{In}(C_k, \mathit{Ls}_i))
\end{split}
\end{equation}

Here, $In(C_k, \mathit{Ls}_i)$ checks whether the line segment, $\mathit{Ls}_i$ is from the cluster $C_k$ or not. In Fig.~\ref{fig:toy_example}, $(x_t, y_t)$ is inside cluster $C_1$ as the imaginary line paralleled to axis $x$ from it intersects an odd number (three) of line segments of the cluster. On the other hand, the imaginary line intersects two line segments of cluster $C_2$, and thus, we can conclude that, the data point is outside of the cluster $C_2$. 
\end{itemize}

For formal modelling, let us assume that a total of $n_s$ different sensor measurements are captured from the body of a person, and he/she possesses a health state from the $n_l$ labels. Without the loss of generality, we assume that each measurement is recorded/reported by one sensor. Let $\mathcal{S}$ be the set of all the sensors,
$\mathcal{P}$ be the set of measurements of those sensors, and $\mathcal{L}$ be the set of all possible labels for a person.

Besides, let the current label of the person be $j$, where $j \in \mathcal{L}$. To keep the clustering simple (2D) and so the associated constraints, we consider the relationship of two sensors at a time. Thus, for the label $j$ $\in \mathcal{L}$ and sensor pair ($a$, $b$) where ($a$, $b$) $\in$  $\mathcal{S}$ we get one or more clusters, $\mathcal{C}_{k}^{a, b, j}$, representing the relationship between the two measurements for that specific label. These clusters consist of a few line segments, which are represented as $\mathit{Ls}_l^{a, b, j}$.  To check the consistency of the data measurements with the constraints from DBSCAN, let us take pair of two sensor measurements $\mathcal{P}_{a,b}$ = $(\mathcal{P}_{a}, \mathcal{P}_{b})$ $\in \mathcal{P}$. We verify the consistency of measurement set $\mathcal{P}$ by checking if each pair of measurements is within any of the corresponding clusters, $\mathcal{C}_k^{a, b, j}$. The measurement set is consistent if the following condition holds: 
%
\begin{equation}\label{within_cluster}
    \begin{split}
     & \mathit{consistent}(\mathcal{P},j) \iff \\ & ~~~\forall_{(a,b) \in \mathcal{S} \land (a!=b)} \exists_{k \in \mathcal{C}}
     \mathit{withinCluster}(\mathcal{P}_{a,b},\mathcal{C}_k^{{a, b,j}})
\end{split}
\end{equation}

where,
\begin{equation}
 \begin{split}
  \mathit{withinCluster}&(\mathcal{P}_{a,b},\mathcal{C}_k^{{a, b, j}}) \iff \\
  \bigoplus_{l}  &(\mathit{intersect}(\mathcal{P}_{a}, \mathcal{P}_{b}, \mathit{Ls}_l^{a, b, j}) 
    \land In(\mathcal{C}_k^{a, b, j}, \mathit{Ls}_l^{a, b, j}))
\end{split}
\end{equation}
and, 
\begin{equation}
 \label{split_0}
\begin{split}
\mathit{intersect}&(\mathcal{P}_{a}, \mathcal{P}_{b}, Ls_l^{a, b, j}) \iff \\ 
& \mathit{inRangeOfLineSegment}(\mathcal{P}_{a}, \mathcal{P}_{b}, \mathit{Ls}_l^{a,b, j}) \\ 
& \land \mathit{leftOfLineSegment}(\mathcal{P}_{a}, \mathcal{P}_{b}), \mathit{Ls}_l^{a,b,j})
\end{split}
\end{equation}

\noindent\textbf{K-means Algorithm-Based constraints:}
K-means algorithm-based constraints acquisition requires a similar approach as constraint acquisition of the DBSCAN algorithm. Due to the algorithmic variation, the number of clusters, number of noise points, and clusters' points are different.

\vspace{12pt}
\noindent{\slshape \bf Formal Modeling of Attack Model:}

\noindent\textbf{Formal Modeling of Attacks:}
Formal models let us explore the search space of all possible behaviors of the system and figuring out potential vulnerabilities. In our case, the attack model takes the attacker's goal, attacker's capability and, the underlying model of the system as input formally models them and finds out possible threats using the satisfiability modulo theory (SMT). Our proposed framework aims to achieve the attacker's goal compromising a minimum number of sensors within the attacker's capability as well. The attacker's capability is dependent on the range of values that can be changed without alarming the system.

\noindent\textbf{Attacker's Goal:}
In our framework, we consider an adversary attempts to compromise the system so that the control system misclassifies the patient and supply incorrect medication. The framework comes up with all possible attack vectors. Attack vector reveals the complete path of launching an attack. An attacker can change a patient's label from $j$ to $\bar{j}$ if the following constraint is satisfied.
\begin{equation}
\label{eqn:attack_goal}
\begin{split}
alter(j,\bar{j}) \rightarrow \forall_{s\in \mathcal{S}} (\bar{\mathcal{P}_{s}} \rightarrow \mathcal{P}_{s} + \Delta \mathcal{P}_{s})  \land \mathit{inference}(\mathcal{P},j) \\ 
\land \mathit{consistent}(\mathcal{P},j) \land  \mathit{inference}(\bar{\mathcal{P}},\bar{j}) \land \mathit{consistent}(\bar{\mathcal{P}},\bar{j})
\end{split}
\end{equation}

Equation~\ref{eqn:attack_goal} requires both the current and altered labels to be satisfied by {the classification model and ADM constraints.} {Here $\mathcal{P}_{s}$, $\Delta \mathcal{P}_{s}$, and $\bar{\mathcal{P}_{s}}$ represents the actual, amount of change, and altered measurement value of sensor $s$.}

\noindent\textbf{Attacker's Accessibility:}
An attacker can change a sensor measurement if he/she has access to that particular sensor.
\begin{equation}
\label{eqn:attack_const}
\begin{split}
\forall_{s\in \mathcal{S}}~a_s \rightarrow (\Delta \mathcal{P}_s\neq0) 
\end{split}
\end{equation}

Here $a_s$ is converted to integer, where $false$ is replaced by 0 and $true$ is transformed into 1.

\noindent\textbf{Attacker's Capability:}
An attacker may not be able to launch a successful attack by modifying one sensor value only. Our formal model specifies how much resources can be accessed and tempered by the attacker.
\begin{equation}
\begin{split}
\sum_{s\in \mathcal{S}}~a_s \leq Maxsensors
\end{split}
\end{equation}

\begin{equation}
\begin{split}
\forall_{s\in \mathcal{S}}~abs(\frac{\Delta{\mathcal{P}_{s}}}{\mathcal{P}_{s}})<Threshold
\end{split}
\end{equation}

Here $Maxsensors$ limits the maximum number of sensors accessible by the attacker, and  $Threshold$ denotes the allowed range of measurement alteration for achieving the attack goal without getting revealed. 
Even if compromised sensor measurements of a successful attack come from an existing cluster following all ADM constraints, drastic alteration (compared to rcent values) may create suspicion to the controller. Thresholds eare considered for ensuring a stealthy attack.
\noindent\textbf{Attack Matrix:}
Our proposed framework can generate an attack matrix that associates the complete attack vector with all possible attack goals from a specific label that can be launched with the minimal capability of the attacker. Attack Matrix also expresses whether it is feasible to attain a certain goal.

\begin{table}[!t]
\scriptsize
\caption{Attack scenario under different attack model}
\begin{tabular}{|p{1.95 cm}|p{1cm}|p{1cm}|p{1cm}|p{1cm}|}
\hline
                          & \multicolumn{1}{l|}{Heart Rate} & \multicolumn{1}{l|}{Systolic} & \multicolumn{1}{l|}{Diastolic} & \multicolumn{1}{l|}{Blood Oxygen} \\ \hline
Actual measurements             & 122.94                          & 75.98                         & 153.56                         & 93.2                              \\ \hline
Attacked measurements (DT constraints only)            & 122.94                          & 73.46                         & 153.56                         & 98.5                              \\ \hline
Attacked measurements (both DT and DBScan constraints) & 120.332                         & 81.855                        & 149.67                         & 98.5                               \\ \hline
\end{tabular}
\label{tab:case_study}
\end{table}

\begin{table}[!t]
\caption{Example Decision Tree Constraints}
\label{tab:dt_constraints}
\scriptsize
\begin{tabular}{|p{3.3in}|}
\hline
\begin{equation}\nonumber
\begin{split}
inference(\mathcal{P},0) \rightarrow ((\mathcal{P}_5 \leq 20.5) \land (((\mathcal{P}_7\leq 8.5) \land (\mathcal{P}_4\leq 94.005)) \\
\lor ((\mathcal{P}_7\leq 8.5) \land (\mathcal{P}_2 \leq 140.46) \land  (\mathcal{P}_0 \leq 100.51)))) \lor~~ \\
 ((\mathcal{P}_5 > 20.5) \land (\mathcal{P}_3 >130.495)))  
 \end{split}
\end{equation}
\begin{equation}\nonumber
inference(\mathcal{P},1) \rightarrow ((\mathcal{P}_5 >20.5) \land (\mathcal{P}_3 \leq 130.495) \land (\mathcal{P}_2 \leq 140.46)) 
\end{equation}
\begin{equation}\nonumber
inference(\mathcal{P},2) \rightarrow ((\mathcal{P}_5>20.5) \land (\mathcal{P}_3 \leq 130.495) \land (\mathcal{P}_2 >140.46))
\end{equation}
\begin{equation}\nonumber
inference(\mathcal{P},3) \rightarrow (\mathcal{P}_5 \leq 20.5) \land (\mathcal{P}_7>8.5) \land (\mathcal{P}_5\leq20.5)
\end{equation}
\begin{equation}\nonumber
\begin{split}
inference(\mathcal{P},4) \rightarrow (\mathcal{P}_5 \leq 20.5) \land (\mathcal{P}^7 \leq 8.5) \land (\mathcal{P}_4 > 94.005) \land \\ 
(\mathcal{P}_2 \leq 140.46))  \land (\mathcal{P}_0 > 100.51) 
\end{split}
\end{equation}

\begin{equation}\nonumber
\begin{split}
inference(\mathcal{P},5) \rightarrow (\mathcal{P}_5 \leq 20.5) \land (\mathcal{P}_7 \leq 8.5) \land (\mathcal{P}_4 > 94.005) \land\\ (\mathcal{P}_2 > 140.46)
\end{split}
\end{equation}
\\\hline
\end{tabular}
\end{table}

\begin{table}[!t]
\caption{Example DBSCAN algorithm-driven constraints}
\label{tab:dbscan_constraints}
\scriptsize
\begin{center}
\begin{tabular}{|p{0.96\columnwidth}|}
\hline \\
$
... (((-0.75\mathcal{P}_0^0 +0.35\mathcal{P}_1^0 + 79.929 \geq 0) \land (70.11 < \mathcal{P}_1^0 \leq 70.86)) \oplus ((-1.5\mathcal{P}_0^0 +0.53\mathcal{P}_1^0 + 171.7767 \geq 0) \land (68.61 < \mathcal{P}_1^0 \leq 70.11)) \oplus ((-0.08\mathcal{P}_0^0 +0.53\mathcal{P}_1^0 + 171.7767 \geq 0) \land (65.21 < \mathcal{P}_1^0 \leq 67.44)) \lor ((0.08\mathcal{P}_0^0 +0.39\mathcal{P}_1^0 + 1059.126 \geq 0) \land (85.27 < \mathcal{P}_1^0 \leq 87.36)) \oplus ((0.04\mathcal{P}_0^0 +0.43\mathcal{P}_1^0 + 21.929 \geq 0) \land (88.22 < \mathcal{P}_1^0 \leq 92.33)) \oplus  ((0.04\mathcal{P}_0^0 +0.41\mathcal{P}_1^0 + 21.21 \geq 0) \land (87.35 < \mathcal{P}_1^0 \leq 91.15)) )
$
\vspace{3pt}
\\\hline
\end{tabular}
\end{center}
\end{table}

\subsection{Example Case Studies}  \label{case_study}  
In our threat analysis model, we presume that the attacker has complete knowledge of SHS architecture and inter-sensor dependencies and also {has access to the points for compromising data.} The attacker can alter sensor measurements to deliver incorrect medication by compromising the underlying DCM. We assume that controllers and actuators are robust enough and are physically protected; hence, they cannot be directly compromised. Moreover, as the SHS IoMT network sensors cannot consume a significant amount of energy, they are unable to send or receive encrypted data due to the limitation of computation power. An attacker can exploit this vulnerability and launch an attack to compromise the consistency of the system. {We have conducted two case studies for evaluating our framework.} {In both of the presented case studies, we consider an SHS with DT-based DCM and DBSCAN-based ADM as these two ML algorithms performed better than the other considered algorithms based on our experimentation.}

\begin{table}[!t]
\scriptsize
\caption{Attack scenario under different attack model}
\begin{tabular}{|p{1.95 cm}|p{1cm}|p{1cm}|p{1cm}|p{1cm}|}
\hline
                          & \multicolumn{1}{l|}{Heart Rate} & \multicolumn{1}{l|}{Systolic} & \multicolumn{1}{l|}{Diastolic} & \multicolumn{1}{l|}{Blood Oxygen} \\ \hline
Actual measurements             & 122.94                          & 75.98                         & 153.56                         & 93.2                              \\ \hline
Attacked measurements (DT constraints only)            & 122.94                          & 73.46                         & 153.56                         & 98.5                              \\ \hline
Attacked measurements (both DT and DBScan constraints) & 120.332                         & 81.855                        & 149.67                         & 98.5                               \\ \hline
\end{tabular}
\label{tab:case_study}
\end{table}

\subsubsection{Case Study 1}
To verify SHChecker, we develop a realistic healthcare dataset with eight sensor measurements with 17,000 samples after preprocessing. In our processed dataset, we have a total of 6 labels of various patient status.
Table~\ref{tab:case_study} shows the measurements of four sensors of a particular patient. For misclassifying a high blood cholesterol labeled patient into a high blood pressure one, compromising only one sensor can not achieve the attack goal. However, the attacker can be successful by compromising two sensors (e.g., systolic blood pressure and blood oxygen) if we consider only the DT model for threat modeling. However, our combined model does not comply with this. The combined DT and DBSCAN cluster-based algorithm shows that we need to compromise at least four sensors to achieve the goal, which confirms the ADM's necessity in the smart healthcare system. Using the combined model, attack vector can be found by changing the heart rate sensor value by 2.12\%, systolic blood pressure by 7.17\%, diastolic blood pressure value by 2.53\%, and blood oxygen measurement value by 5.68\%. Tables~\ref{tab:dt_constraints} and~\ref{tab:dbscan_constraints} demonstrate some example DT and DBSCAN constraints obtained from the formal threat model.

\begin{table}[!b]
\caption{Example Decision Tree Constraints}
\label{tab:dt_constraints}
\scriptsize
\begin{tabular}{|p{3.3in}|}
\hline
\begin{equation}\nonumber
\begin{split}
inference(\mathcal{P},0) \rightarrow ((\mathcal{P}_5 \leq 20.5) \land (((\mathcal{P}_7\leq 8.5) \land (\mathcal{P}_4\leq 94.005)) \\
\lor ((\mathcal{P}_7\leq 8.5) \land (\mathcal{P}_2 \leq 140.46) \land  (\mathcal{P}_0 \leq 100.51)))) \lor~~ \\
 ((\mathcal{P}_5 > 20.5) \land (\mathcal{P}_3 >130.495)))  
 \end{split}
\end{equation}
\begin{equation}\nonumber
inference(\mathcal{P},1) \rightarrow ((\mathcal{P}_5 >20.5) \land (\mathcal{P}_3 \leq 130.495) \land (\mathcal{P}_2 \leq 140.46)) 
\end{equation}
\begin{equation}\nonumber
inference(\mathcal{P},2) \rightarrow ((\mathcal{P}_5>20.5) \land (\mathcal{P}_3 \leq 130.495) \land (\mathcal{P}_2 >140.46))
\end{equation}
\begin{equation}\nonumber
inference(\mathcal{P},3) \rightarrow (\mathcal{P}_5 \leq 20.5) \land (\mathcal{P}_7>8.5) \land (\mathcal{P}_5\leq20.5)
\end{equation}
\begin{equation}\nonumber
\begin{split}
inference(\mathcal{P},4) \rightarrow (\mathcal{P}_5 \leq 20.5) \land (\mathcal{P}^7 \leq 8.5) \land (\mathcal{P}_4 > 94.005) \land \\ 
(\mathcal{P}_2 \leq 140.46))  \land (\mathcal{P}_0 > 100.51) 
\end{split}
\end{equation}

\begin{equation}\nonumber
\begin{split}
inference(\mathcal{P},5) \rightarrow (\mathcal{P}_5 \leq 20.5) \land (\mathcal{P}_7 \leq 8.5) \land (\mathcal{P}_4 > 94.005) \land\\ (\mathcal{P}_2 > 140.46)
\end{split}
\end{equation}
\\\hline
\end{tabular}
\end{table}

\subsubsection{Case Study 2}
For verifying our model, we have implemented our SHChecker framework in a real-world dataset collected by The University of Queensland~\cite{liu2012university}. The dataset contains 49 sensor measurements of 32 anesthesia patients with surgical cases at the Royal Adelaide Hospital, monitored using Philips intellivue monitors and Datex-Ohmeda anesthesia machine. After preprocessing, the dataset containing 209,115 samples with 26 sensor measurements were used to find attack vectors using our proposed framework. Based on the vital signs of the patients, the monitoring systems issue single or multiple alarms. In our processed dataset, we have a total of 58 labels having 28 different alarms. Out of these labels, 24 deals with a single alarm, 19 handles a couple of alarms, 12 of them provide triple alarms, and the rest involves quadruple alarms. Using our threat model, we checked the feasibility of compromising the system by providing the wrong alarm. Our experimentation shows that by altering 9.2\%, 8.1\%, 8.4\%, 2.3\%, 5.9\%, 9.9\%, 2.1\%, 3.9\% values in artery diastolic pressure (ART Dia), artery mean pressure (ART Mean), effective end-tidal decreased hemoglobin oxygen saturation (ETDES) label, inspired decreased hemoglobin oxygen saturation (INDES) label, end-Tidal isoelectric (ETISO) point, inspired isoelectric (INISO) point, effective end-tidal concentration of sevoflurane  (ETSEV) and inspired concentration of sevoflurane (INSEV) sensors respectively, an adversary can make the system trigger APNEA, low blood pressure (NBPsLOW), low end-tidal carbon-dioxide (etCO2LOW), high inspired concentration of sevoflurane (inSEVHIGH) alarm instead of APNEA, high minute volume (MVexpHIGH) by compromising both DT and DBSCAN-based control algorithms.


\label{sec:technical_details}

\section{Implementation}
\label{sec:implementation}
\begin{figure*}[t]
    \begin{center}
        \subfigure[]
        {
        \label{patient_monitor}
            \includegraphics[width=1.15\columnwidth]{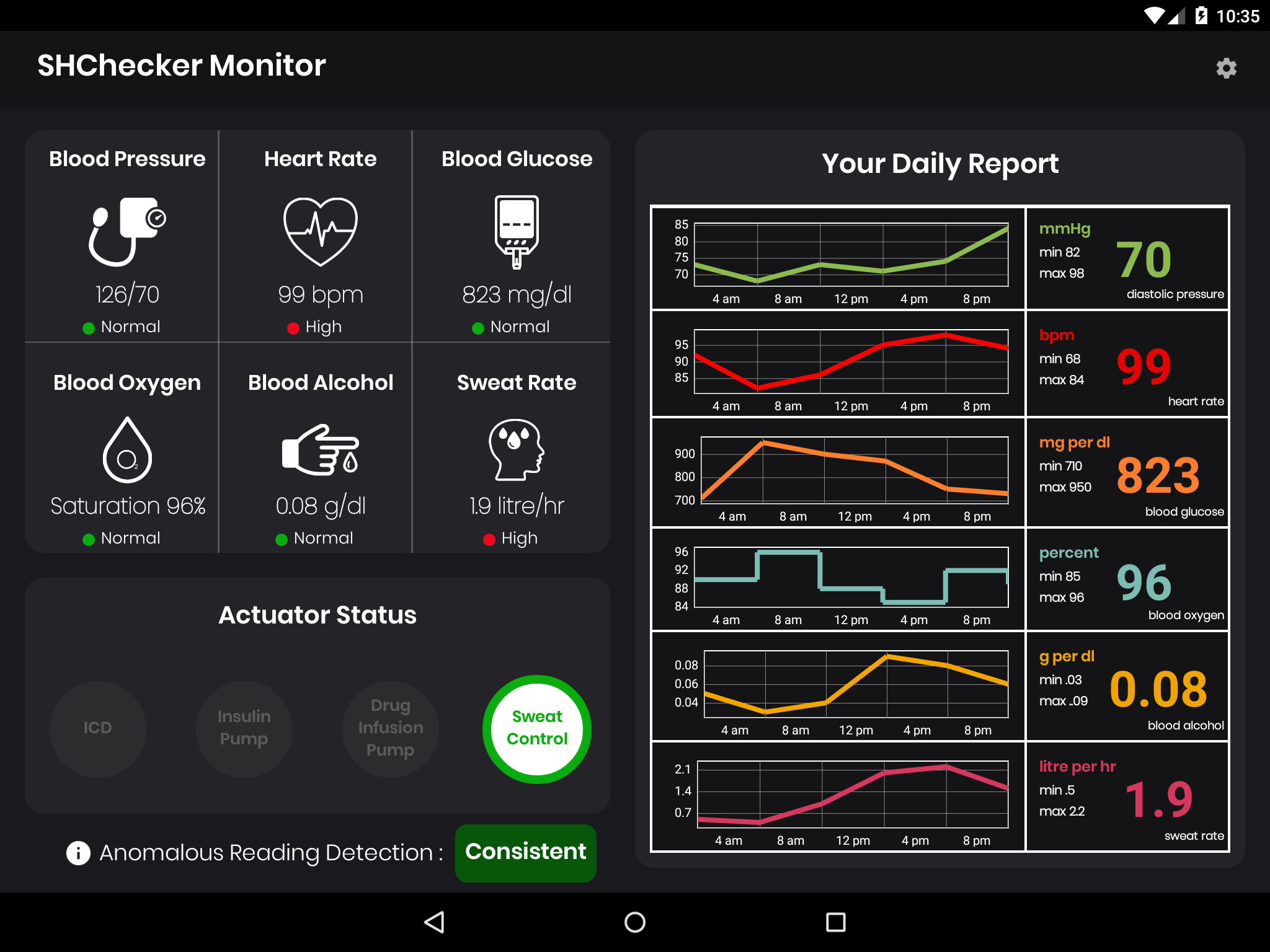}
        }
        \subfigure[]
         {
        \label{firebase}
            \includegraphics[width=0.75\columnwidth]{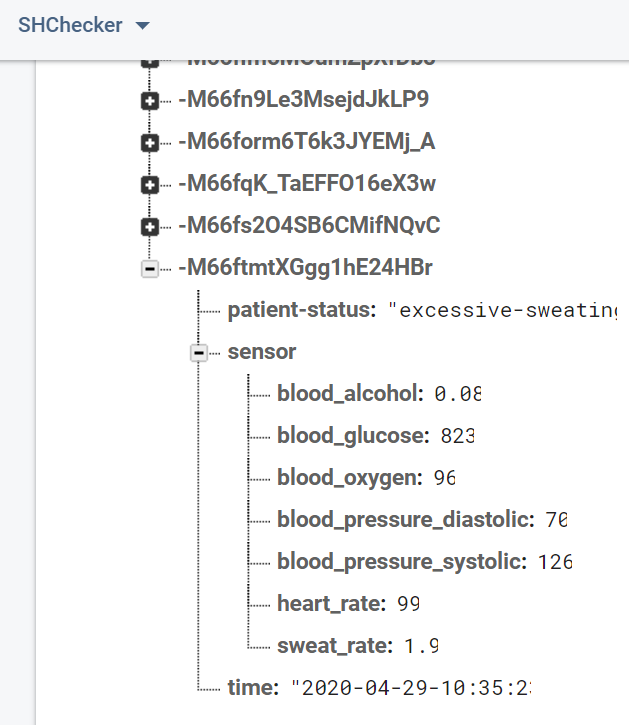}
        }
    \end{center}
    \caption{(a) Patient monitor (b) snapshot of database.}
    \label{fig:snapshot}
\end{figure*}

\begin{figure}[!htb]
    \centering
    \includegraphics[width=1\columnwidth]{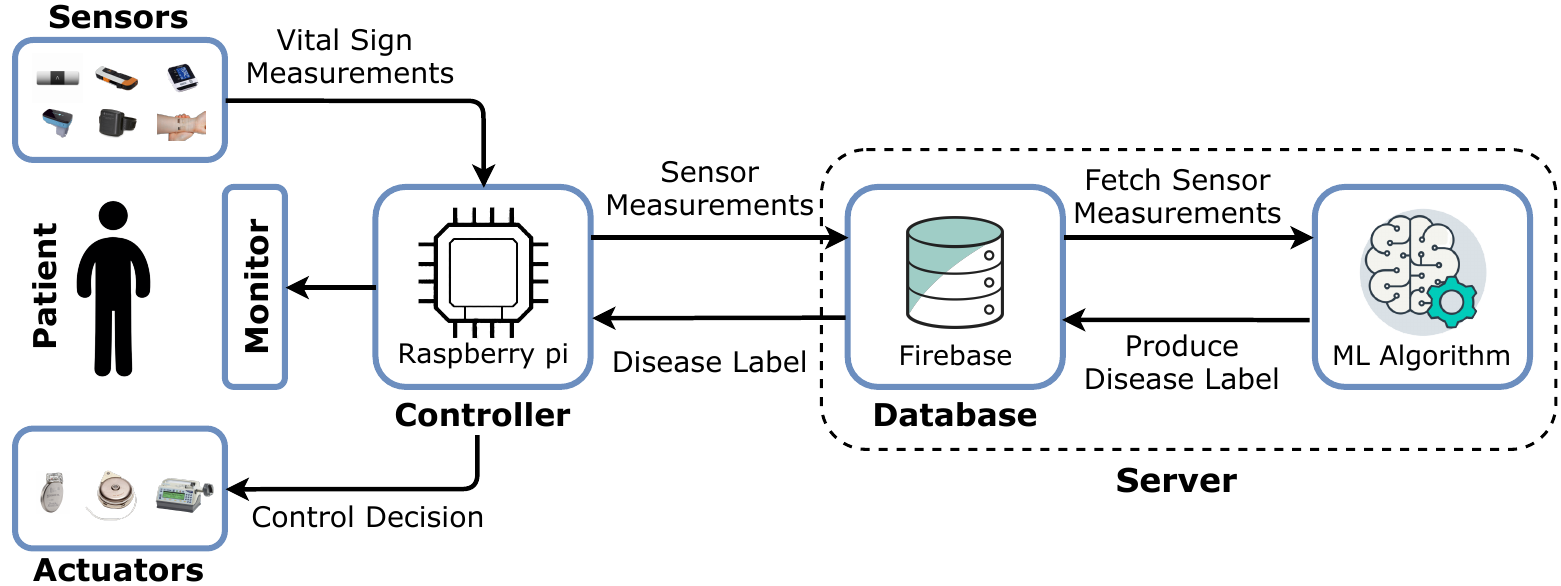}
    \caption{Testbed implementation of an SHS}
    \label{fig:testbed}
\end{figure}

For verifying our framework, We implement an SHS testbed considering a patient being connected to an IoMT network and is being continuously monitored with six different body sensors and is getting real-time treatment with four actuators. Fig.~\ref{fig:testbed} shows our testbed setup where the sensors and the actuators are connected with a raspberry pi-based controller. The controller receives the sensor values by maintaining a specific time interval as the sampling rate of the sensors is different and acquires the disease label from the machine learning algorithms running inside the server. The controller and the server exchange data using a real-time (firebase) NoSQL database as it is difficult to handle data generated by the IoT system with traditional databases~\cite{singh2019iot}. We develop an android application that works as a patient monitor showing sensor measurements and the IMD devices' status. In the monitor, we show the sensor values and the actuators triggered by the controller. Fig.~\ref{fig:snapshot}(a) displays a snapshot of our android application for patient monitoring, and Fig.~\ref{fig:snapshot}(b) exhibits NoSQL data structure from the firebase database for storing patients vital sign data.

\label{sec:implementation}

\section{Evaluation}
\label{sec:evaluation}
We consider the following set of research questions to evaluate our proposed threat framework.

\textbf{RQ1} What are the impacts of using a disease classification and anomaly detection-based combined model instead of using a standalone model? (Section \ref{combined_model})

\textbf{RQ2} What is the impact of using a combination of all pairs of features instead of using all features together? (Section~\ref{pair})

\textbf{RQ3} What is the performance of the system varying attacker's capability? (Section~\ref{attacker_capability})

\textbf{RQ4} What are the most significant devices that need to be compromised while launching an attack? (Section~\ref{significant_device})


\textbf{RQ5} What is the resiliency of the system? (Section~\ref{resiliency})

\textbf{RQ6} What is the impact of execution time for an increasing number of sensors? (Section~\ref{scalability})

\subsection{Environmental Setup and Methodology} \label{Sec:setup_env}
We leverage  SMT~\cite{barrett2018satisfiability}, which determines whether a first-order formula is achievable with respect to some logical theory. We encode them using boolean and real terms. Based on the satisfiability,  the model produces SAT or UNSAT results. The SAT result, in our case, indicates an attack vector, the complete information about launching a successful attack on the SHS. For the UNSAT case, there is no possible attack vector under those constraints.

{The formal threat modeling has been implemented using Z3 library~\cite{de2008z3}. We have used python API of Z3 SMT solver. } To determine the concave hull boundary for cluster modeling, we used MATLAB boundary function and for evaluating clustering algorithm performance, we have used various R packages. The experiments are conducted on Dell Precision 7920 Tower workstation with Intel Xeon Silver 4110 CPU @3.0GHz, 32 GB memory, 4 GB NVIDIA Quadro P1000 GPU. For evaluating the system, we use a synthetic dataset with eight important vital features of human body, which contains almost 17000 samples and a real dataset titled, The University of Queensland Vital Signs (UQVS) dataset with 209,115 samples measuring 26 vital signs.

\subsection{SHS Performance Metric} \label{performance_measure} 
For evaluating the performance of various ML models for SHS disease classification, we use four different performance metrics: accuracy, precision, recall, and f1-score. Accuracy calculates the number of correctly identified samples of overall data samples. Precision is the measure of false-positive rate, whereas recall quantifies false-negative rate. F1-score takes both precision and recall into account by performing harmonic mean of them. Table~\ref{tab:compare_ml} shows that for both synthetic and generated datasets, DT works better than LR and NN based on accuracy, precision, recall, and f1-score. 

For assessing the performance of ADMs, three different performance metrics have been considered. Table~\ref{tab:compare_cluster} presents a comparative analysis of DBScan and K-means clustering based on some internal cluster validation metrics naming average Silhouette Coefficient Score (SCS), Davies-Bouldin Score (DBS), and Dunn's Index (DI). SCS is a measure of how similar an object is to its cluster (cohesion) compared to other clusters (separation). SCS is a measure of similarity of a cluster sample to its clusters rather than with other clusters. High value of SCS (close to +1) of a cluster specifies its likeness with its cluster where a low value (close to -1) indicates the opposite. In our experiment, we measure SCS of clusters by performing mean of all data points of that particular cluster, defined as the average similarity measure of each cluster with its most similar cluster, where similarity is the ratio of within-cluster distances to between-cluster distances. DBS finds the average similarity measure of each cluster compared to its most similar cluster. The clustering algorithm having lower DBS shows better performance. Again, the DI value assesses compactness and clusters separation measure of the algorithm. Unlike SCS and DBS, higher DI value denotes better performance for clustering algorithms~\cite{datanovia2019}. 
{It is apparent from the analysis that DBSCAN has outperformed K-Means algorithms based on our experimental setup.}

\begin{table}[t]
\centering
\caption{Comparison of the performance of ML algorithms for disease classification} 
\small
\label{tab:compare_ml}
\begin{tabular}{|c|c|c|c|c|}
\hline
\multicolumn{2}{|l|}{} &
  \multicolumn{1}{l|}{\textbf{NN}} &
  \multicolumn{1}{l|}{\textbf{DT}} &
  \multicolumn{1}{l|}{\textbf{LR}} \\ \hline
 & \textbf{Accuracy}  & { 0.91} & { 0.93} & { 0.88} \\
 & \textbf{Precision} & { 0.92} & { 0.92} & { 0.9}  \\
 & \textbf{Recall}    & { 0.89} & { 0.93} & { 0.9}  \\
\multirow{-4}{*}{\textbf{\begin{tabular}[c]{@{}c@{}}Synthetic Data\end{tabular}}} &
  \textbf{F1-Score} &
  { 0.9} &
  { 0.92} &
  { 0.9} \\ \hline
 & \textbf{Accuracy}  & { 0.94} & { 0.97} & { 0.86} \\
 & \textbf{Precision} & { 0.9}  & { 0.92} & { 0.88} \\
 & \textbf{Recall}    & { 0.91} & { 0.92} & { 0.86} \\
\multirow{-4}{*}{\textbf{\begin{tabular}[c]{@{}c@{}}UQVS Data\end{tabular}}} &
  \textbf{F1-Score} &
  { 0.9} &
  { 0.92} &
  { 0.87} \\ \hline
\end{tabular}
\normalsize
\end{table}

\begin{table}[]
\centering
\caption{Comparison of the performance of ML algorithms for anomaly detection}
\scriptsize
\label{tab:compare_cluster}
\begin{tabular}{|p{1cm}|p{.8cm}|p{.8cm}|p{.8cm}|p{.8cm}|p{.8cm}|p{.8cm}|}
\hline
{ } &
  \multicolumn{3}{c|}{{ \textbf{Synthetic Data}}} &
  \multicolumn{3}{c|}{{ \textbf{UQVS Data}}} \\ \cline{2-7} 
\multirow{-2}{*}{{ \textbf{}}} &
  { \textbf{DBS}} &
  { \textbf{SCS}} &
  { \textbf{DI}} &
  { \textbf{DBS}} &
  { \textbf{SCS}} &
  { \textbf{DI}} \\ \hline
{ \textbf{K-means}} &
  { 2.57} &
  { 0.18} &
  { 0.057} &
  { 1.841} &
  { 0.098} &
  { 0.072} \\ \hline
{ \textbf{DBSCAN}} &
  { 0.681} &
  { 0.732} &
  { 0.235} &
  { 0.517} &
  { 0.612} &
  { 0.412} \\ \hline
\end{tabular}
\end{table}

\subsection{Evaluation Results and Discussion} \label{evaluation_results} 

\subsubsection{Evaluation of System Considering a Combined Model} \label{combined_model} 

Due to the sensitivity of data and system models, underlying models of SHS are hardly exposed to public research. 
{However, we consider a IoMT-based SHS with two different types of ML models for threat analysis.} Based on the performance analysis, let us formally model SHS considering DT-based DCM and DBSCAN-based ADM. The DT-based model tends to find a splitting point considering the minimum number of sensor measurements to clearly distinguish a group of vital signs from one patient status to another one. But DT does not consider the inter-relation between all other sensor values for a particular state, which creates a need for an ADM in addition to a DCM.

Consequently, leveraging a clustering algorithm in the SHS model can accumulate the relationship between all sensors for a particular patient state 
{and make the system robust}. 
{Although, NN-based DCMs might capture the inter-relationship between sensor measurements, clustering-based approaches are required for outlier detection as NN always puts a label ignoring the fact of data being an outlier.} 
{Consideration of such model impose constraints on sensor measurement alteration for the adversaries.} An attacker can not alter a patient's status with the knowledge of DT-based constraints only. Because, by compromising sensor values, the attacker could generate a sample that is satisfied by the DT-based model but labeled as an anomaly by the DBSCAN-based model 
{as demonstrated in the case studies}. That's why for analyzing vulnerabilities of a robust healthcare system, our proposed threat model utilizes an ADM along with the actual DCM.

\begin{figure}[!t]
    \begin{center}
        \subfigure[]
        {
        \label{graph_1}
            \includegraphics[width=0.46\columnwidth]{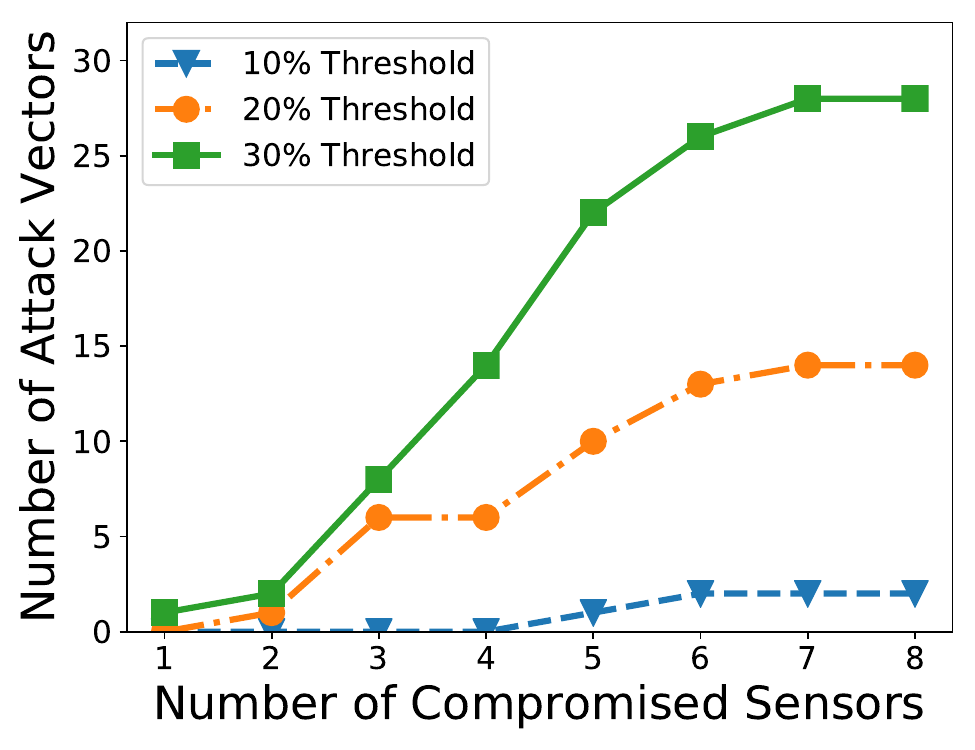}
        }
        \subfigure[]
        {
        \label{real_graph_1}
            \includegraphics[width=0.46\columnwidth]{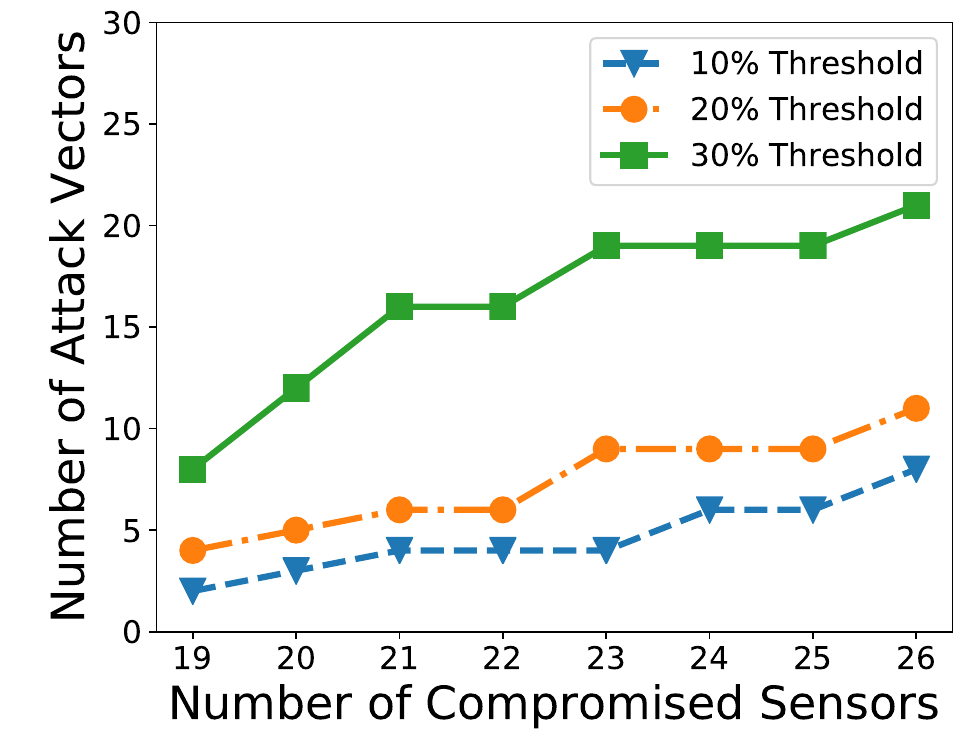}
        }     
    \end{center}
    \caption{SHChecker's performance w.r.t. attacker's capability for {(a) synthetic dataset and (b) UQVS dataset.}}
    \label{fig:av_attcapab}
\end{figure}

\subsubsection{Evaluation of System Considering All Pair of Features} \label{pair}
In our proposed threat framework, we do not consider the relationship among all features for the ADM. Finding relationship among all features together come up with better ADM but over-complicates the constraints 
{increasing solver complexity} and makes the threat analysis infeasible in the case of the automated real-time healthcare system. Besides, to draw a concave hull in the n-th dimensional space, we need at least n-1 points. Many clusters violate this constraint while all feature relationships are considered together. In the case of a pair of features consideration model, some abnormal data from the overall model are labeled as positive data but never encounters any positive data to be anomalous. Table~\ref{tab:rel_mod} shows that pair of feature consideration can capture 95.43\% and 96.64\% anomalies for the synthetic and the UQVS dataset respectively. Considering aforementioned analysis, pair of features consideration model has been adopted in our framework.

\begin{table}[]
\centering
\caption{Performance analysis of pair of relationship model}
\scriptsize

\label{abcdefgh}
\begin{tabular}{|c|p{1.2cm}|p{1.2cm}|p{1.2cm}|p{1cm}|}
\hline
 &
  \multicolumn{2}{c|}{\textbf{Synthetic Dataset}} &
  \multicolumn{2}{c|}{\textbf{UQVS Dataset}} \\ \hline
\textbf{Model} &
  \textbf{\begin{tabular}[c]{@{}c@{}}Positive \\ Sample\end{tabular}} &
  \textbf{\begin{tabular}[c]{@{}c@{}}Noisy \\ Sample\end{tabular}} &
  \textbf{\begin{tabular}[c]{@{}c@{}}Positive \\ Sample\end{tabular}} &
  \textbf{\begin{tabular}[c]{@{}c@{}}Noisy \\ Sample\end{tabular}} \\ \hline
\textbf{\begin{tabular}[c]{@{}c@{}}All measurements \\ consideration\end{tabular}} &
  { 13822} &
  { 3178} &
  { 182,316} &
  { 26,799} \\ \hline
\textbf{\begin{tabular}[c]{@{}c@{}}Combination of pair \\ of measurements \\ consideration\end{tabular}} &
  { 13967} &
  { 3033} &
  { 183,215} &
  { 25,900} \\ \hline
\end{tabular}
\label{tab:rel_mod}
\end{table}

\begin{figure}[!t]
\label{fig:bar_chart}
    \begin{center}
        \subfigure[]
        {
        \label{bar_chart}
            \includegraphics[width=0.7\columnwidth]{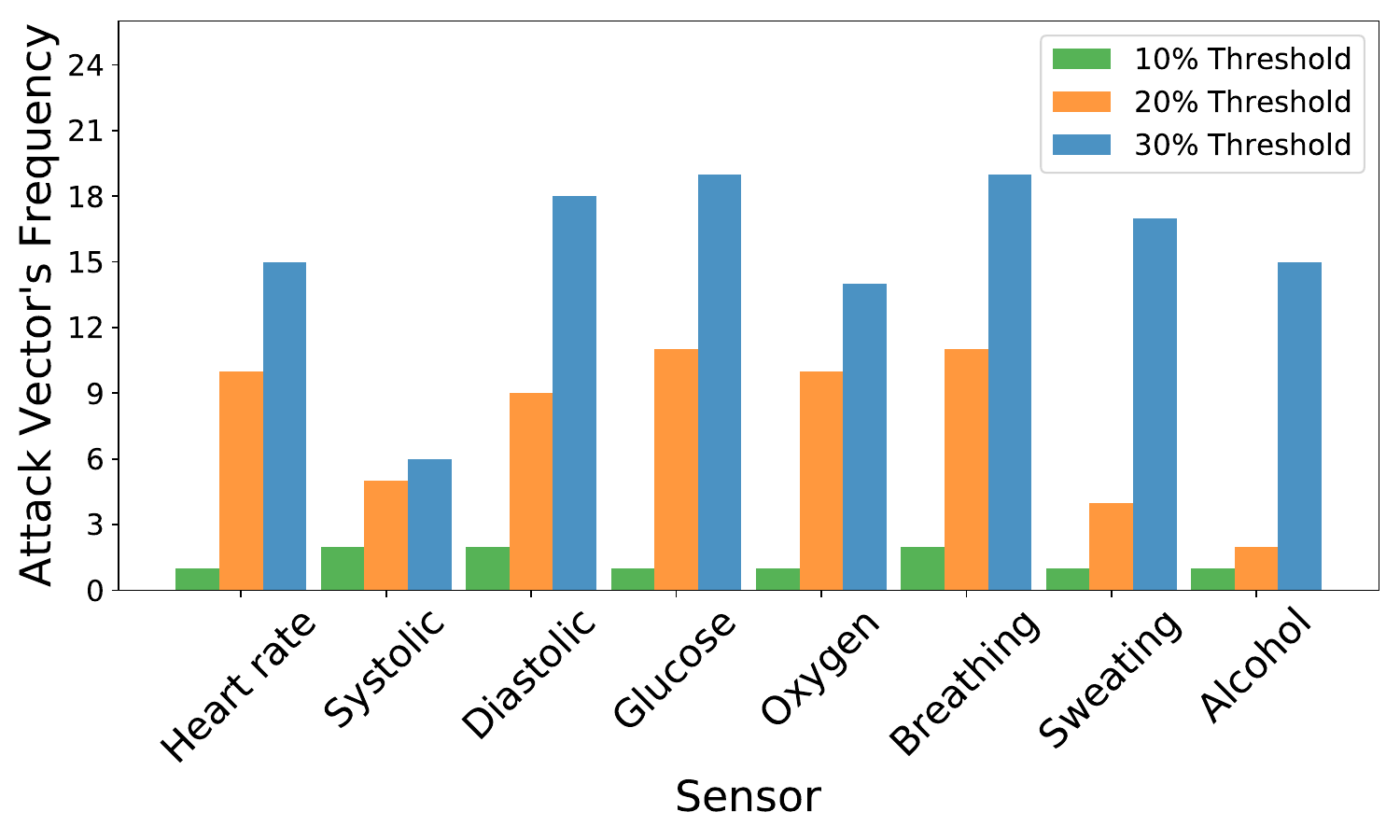}
        }
        \subfigure[]
         {
        \label{real_bar_chart}
            \includegraphics[width=0.7\columnwidth]{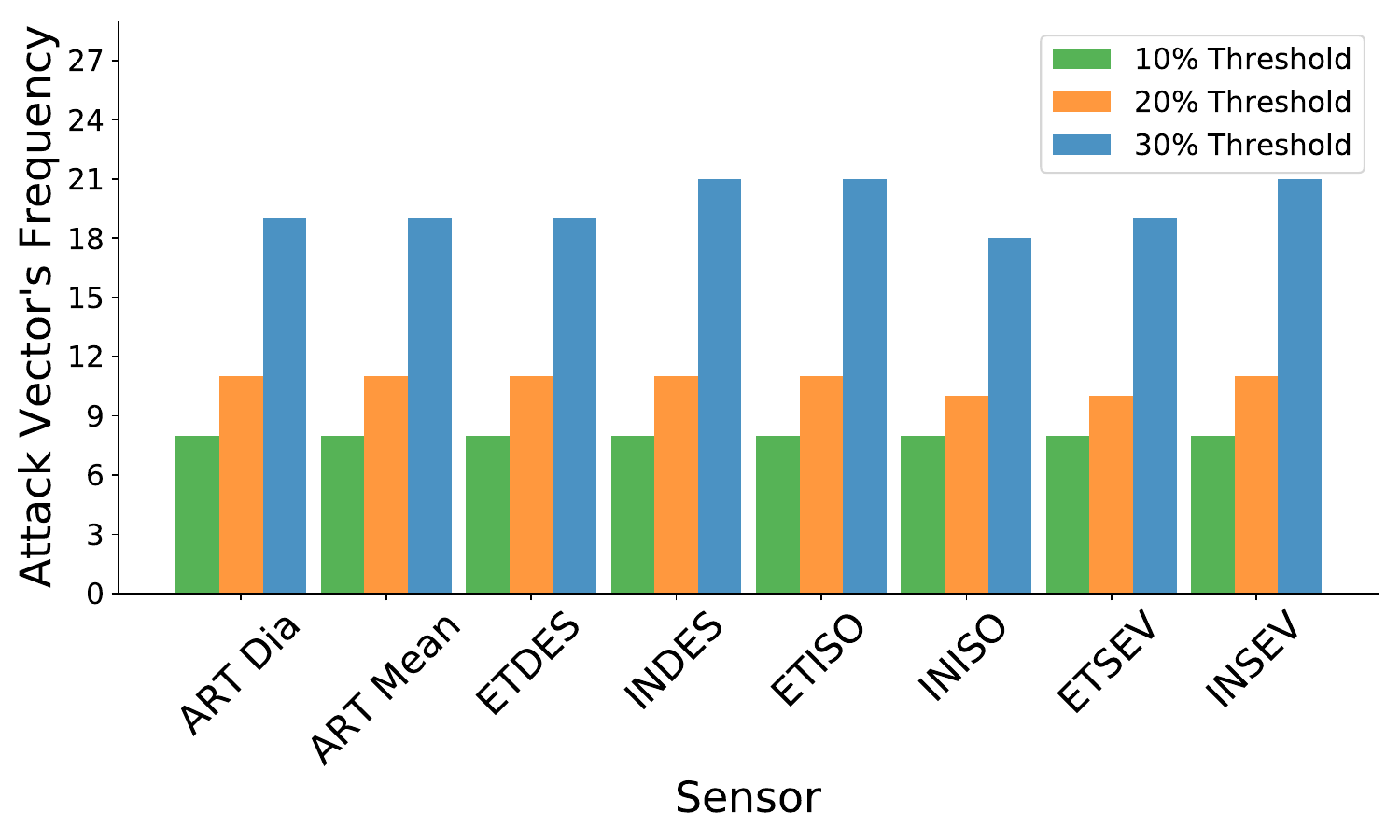}
        }
    \end{center}
    \caption{Frequency of the individual sensors in attack vectors for (a) synthetic data and (b) UQVS data.} 
    \label{fig:bar_chart}
\end{figure}

\begin{figure*}[!htb]
    \begin{center}
        \subfigure[]
        {
        \label{graph_2}
          \includegraphics[width=0.46\columnwidth]{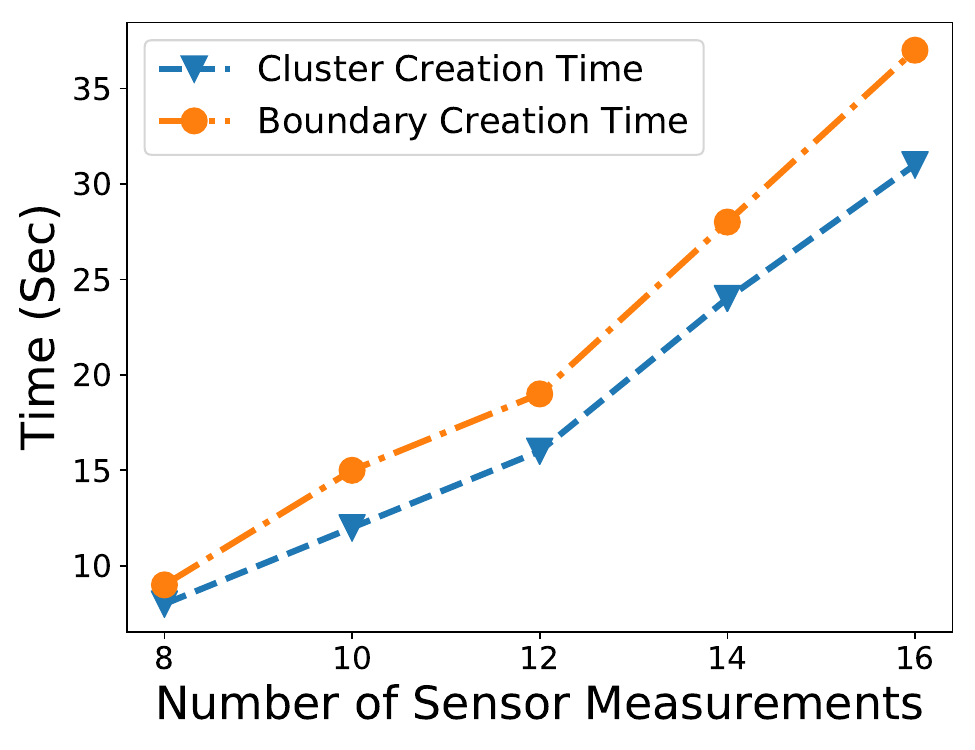}
        }
        \subfigure[]
        {
        \label{graph_3}
            \includegraphics[width=0.46\columnwidth]{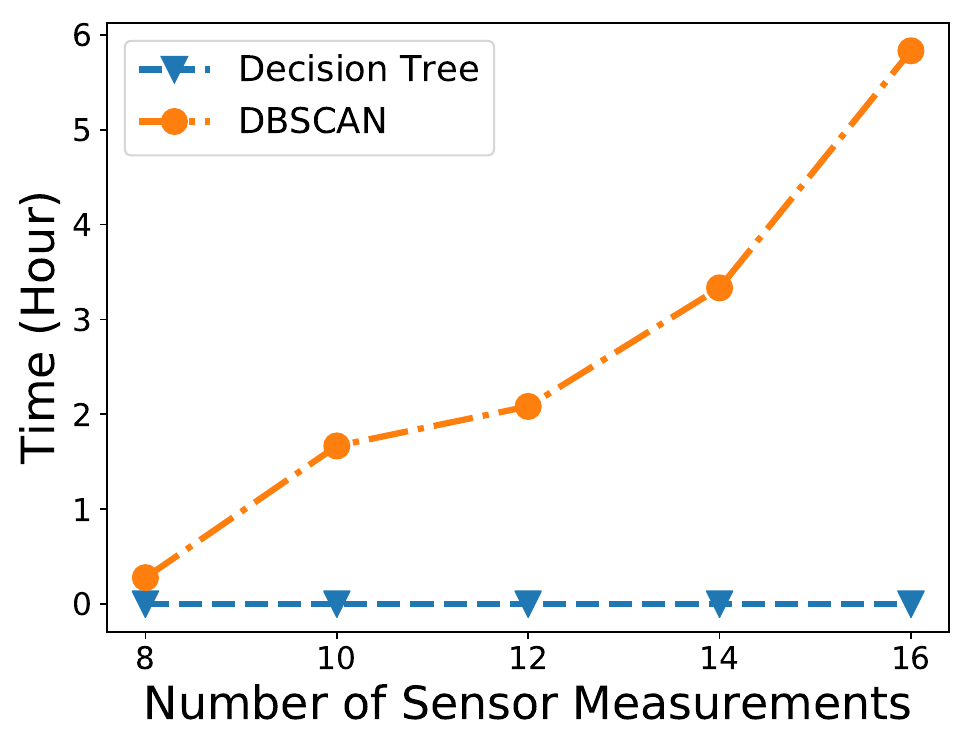}
        }     
        \subfigure[]
        {
        \label{graph_4}
            \includegraphics[width=0.46\columnwidth]{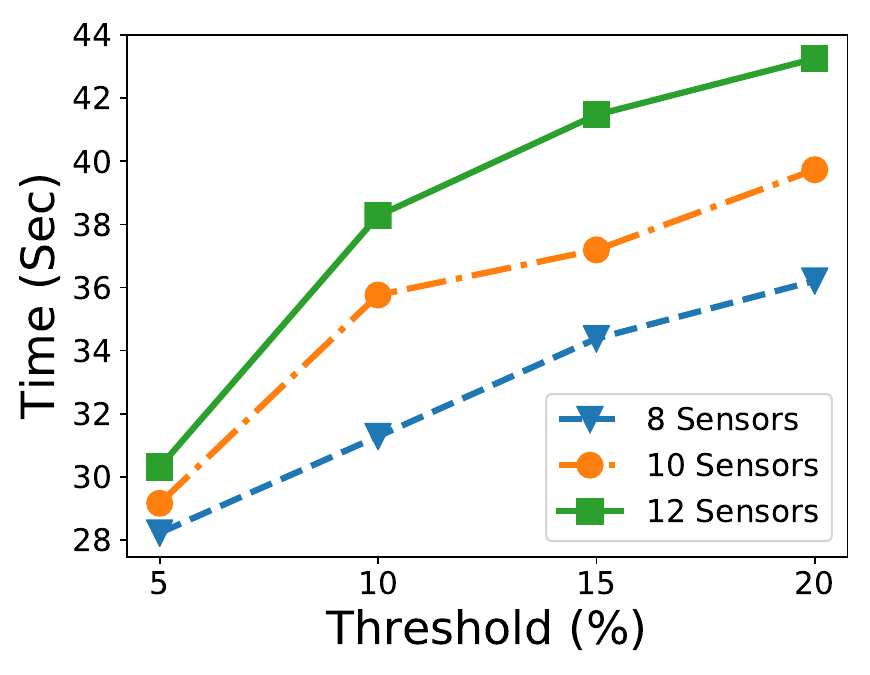}
        }
        \subfigure[]
         {
        \label{graph_5}
            \includegraphics[width=0.46\columnwidth]{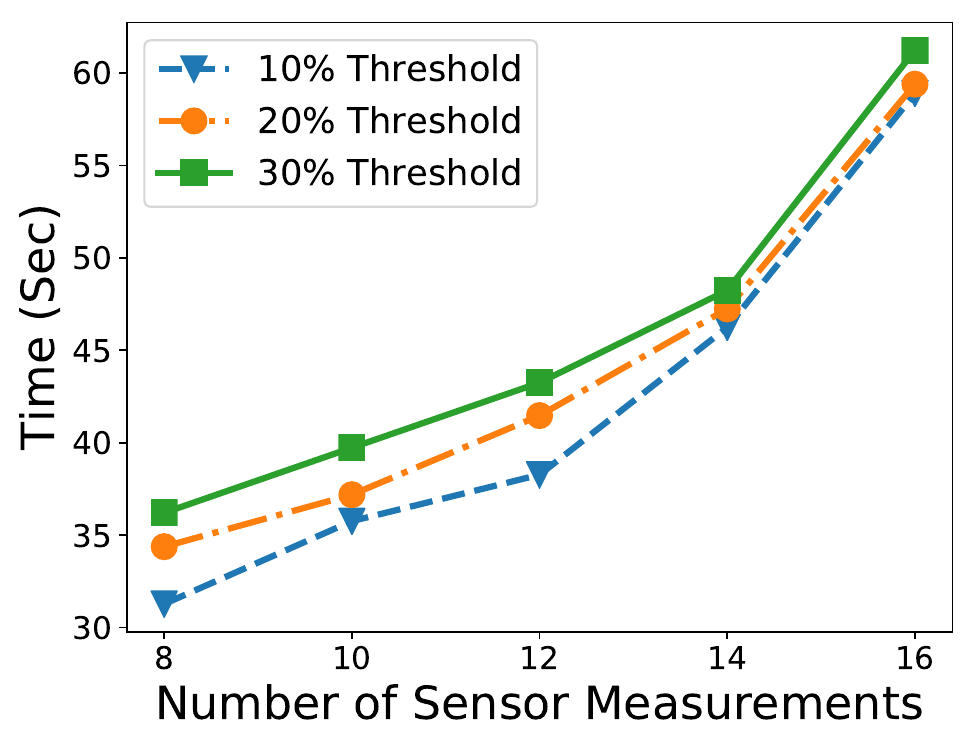}
        }
    \end{center}
    \caption{Execution time of the (a) cluster and boundary creation time based on number of sensor measurements, (b) machine learning constraints generation based on number of sensors, (c) threat analysis based on threshold for data injection, and (d) threat analysis based on the number of sensor measurements measured from the synthetic dataset}
    \label{fig:scalability}
\end{figure*}

\begin{figure*}[!t]
    \begin{center}
        \subfigure[]
        {
        \label{real_graph_2}
            \includegraphics[width=0.46\columnwidth]{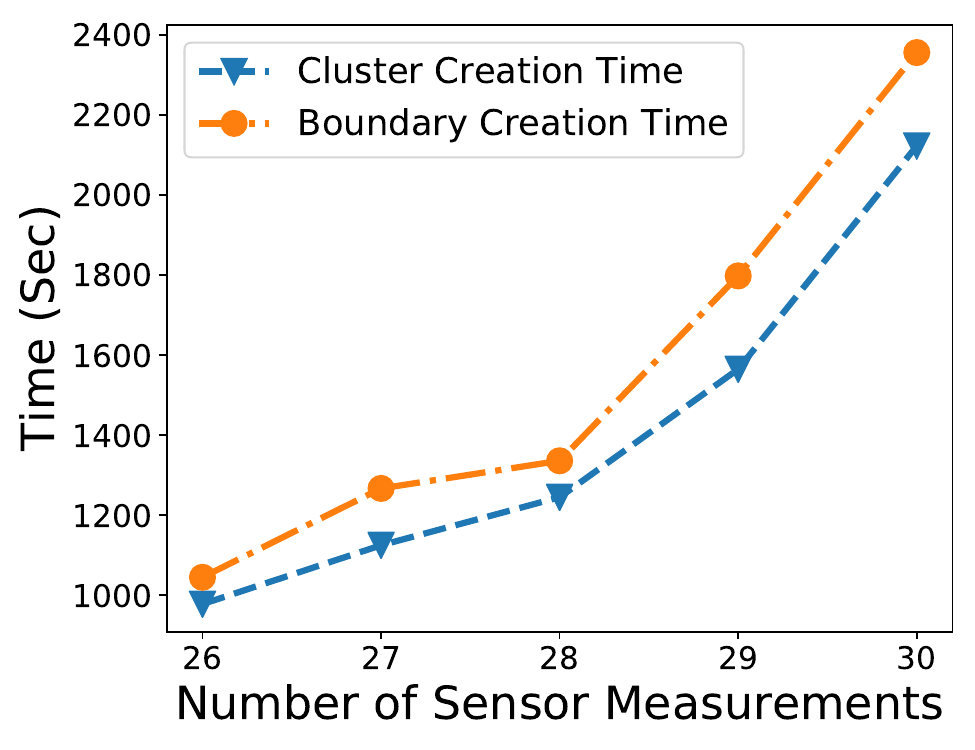}
        }
        \subfigure[]
        {
        \label{real_graph_3}
            \includegraphics[width=0.46\columnwidth]{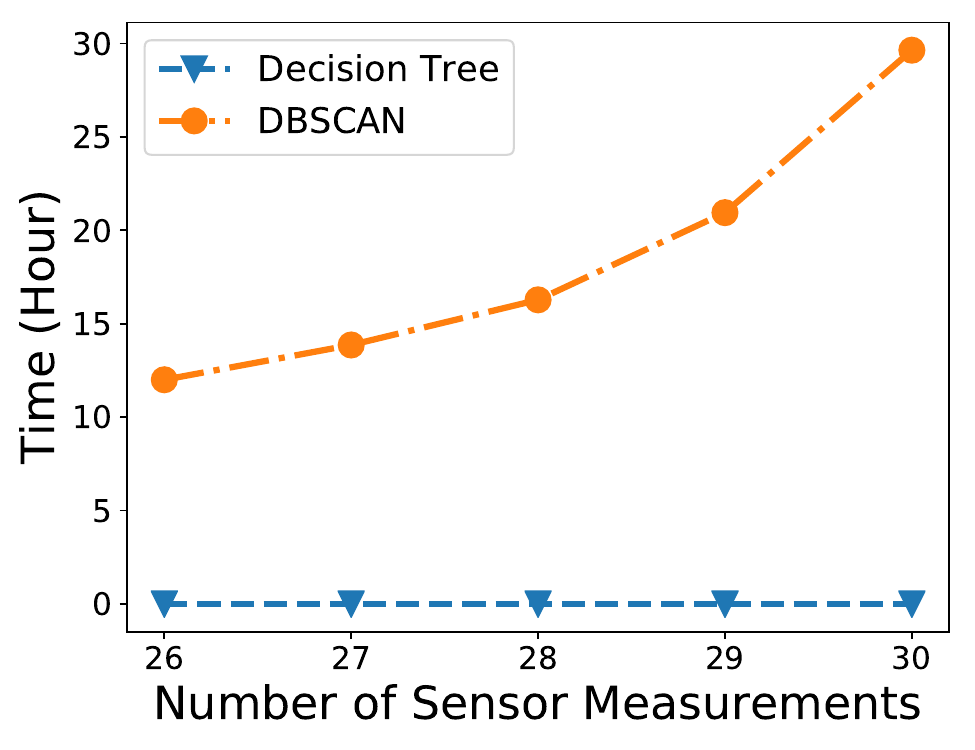}
        }     
        \subfigure[]
        {
        \label{real_graph_4}
            \includegraphics[width=0.46\columnwidth]{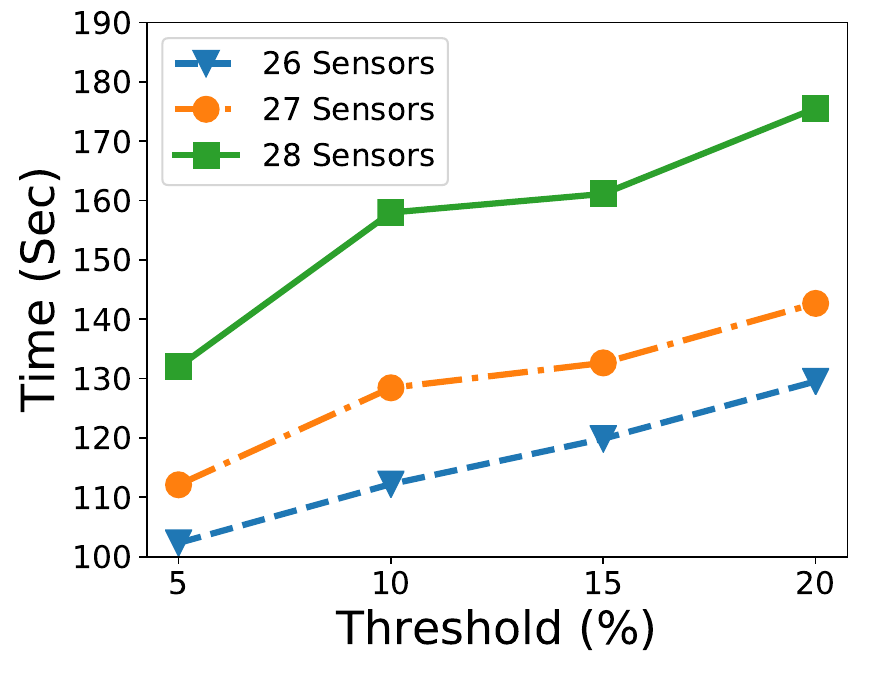}
        }
        \subfigure[]
         {
        \label{real_graph_5}
            \includegraphics[width=0.46\columnwidth]{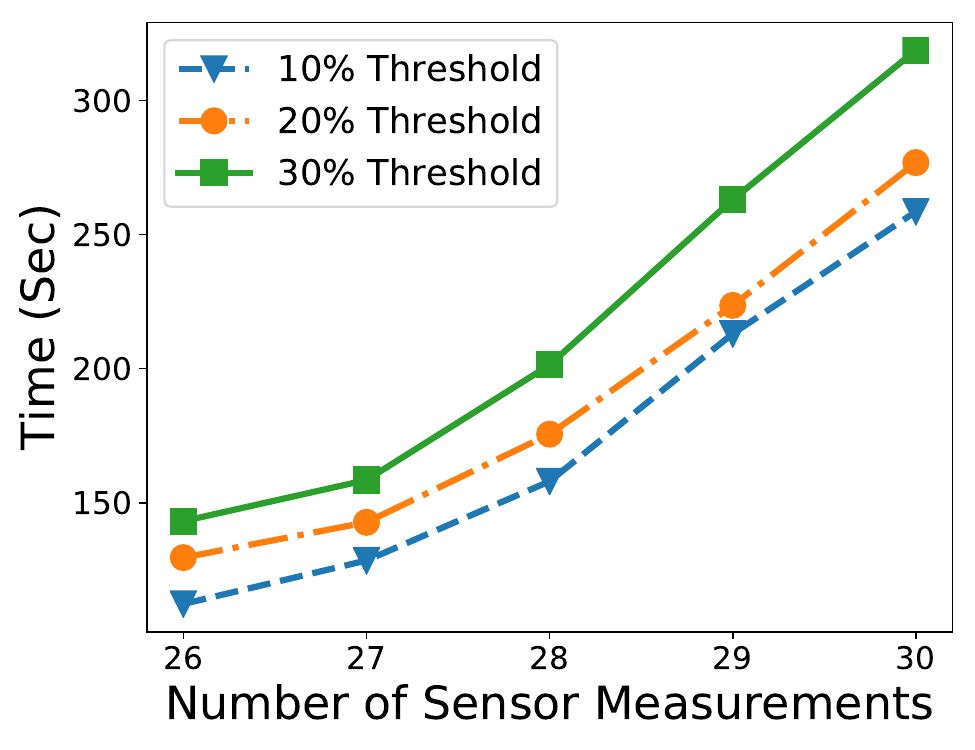}
        }
\end{center}
\caption{Execution time of the (a) cluster and boundary creation time based on number of sensor measurements, (b) machine learning constraints generation based on number of sensors, (c) threat analysis based on a threshold for data injection, and (d) threat analysis based on number of sensor measurements measured from the UQVS dataset.}
\label{fig:scalability_real}
\end{figure*}

\subsubsection{Performance with respect to Attacker's Capability} \label{attacker_capability}

In this part, we evaluate the performance of our proposed SHChecker model by analyzing the total number of attack vectors concerning the attacker's capability. Fig.~\ref{fig:av_attcapab} shows the number of found attack vectors for the different number of compromised sensors as well as the threshold for data injection. The figure shows the attacker is successful in finding an attack vector, even attacking only one sensor. By compromising more sensors, the attacker can achieve a maximum of 3 attack vectors when the injected data is bounded within a threshold of 10\% of the actual sensor data. However, the model finds more attack vectors when the injection threshold is increased. The model is capable of finding 28 different attack vectors, when the attacker is able to attack eight sensors, altering the measurement up to 30\% of the actual value. Although SHChecker considers combination of pair of features, the attack vectors we get using formal threat modeling conform with the actual ML models. We provided the altered sensor measurements obtained from the attack vectors as a input to the DBSCAN and DT algorithms and found the output to be consistent with the attack goals in all cases.

\subsubsection{Frequency of Sensors in The Attack Vectors}
\label{significant_device}
SHChecker framework analyzes all the attack vectors and determines the participation of the individual sensors. We plot Fig.~\ref{fig:bar_chart} to represent the frequency of the sensors in the attack vectors for both datasets. From the figure, we see that all the sensors, except a few (i.e., Sensor 2), participate in attack generation almost equally for 30\% attack threshold. Thus, any of the sensors of the SHS can be compromised to achieve attacker's goal. Besides, this study gives an insight on which sensors should get more attention while developing a defensive tool for the SHSs. 
For example, if frequency of certain sensor measurements in attack vectors is much higher than the others, the sensors associated with the measurements must get extra attention for getting secured. Thus, SHChecker can be a useful tool to provide guidelines for SHS design.

\begin{table}[]
\centering
\caption{Devices to compromise to achieve an attack goal}
\label{resiliency_table}
\scriptsize
\begin{tabular}{|p{1.2cm}|p{1.5cm}|c|p{2.2cm}|}
\hline
\textbf{Current State}  & \textbf{Target State} & {\textbf{\begin{tabular}[c]{@{}c@{}}Number of Sensors \\ to Compromise\end{tabular}}} & \textbf{Sensors}               \\ \hline
Normal                  & High Cholesterol      & 3                                        & Heart rate, Glucose, Alcohol   \\ \hline
High Blood Pressure     & Abnormal Oxygen Level & 4                                        & Systolic, Diastolic, Oxygen, Breathing              \\ \hline
High Cholesterol        & Excessive Sweating    & 2                                        & Heart rate, Breathing \\ \hline
\end{tabular}
\end{table}

\subsubsection{Resiliency Analysis of the System} \label{resiliency}

A system is said to be resilient to the degree to which it rapidly and effectively protects its critical capabilities from the disruption caused by adverse events and conditions. Our proposed threat model can determine the resiliency of a system for a particular attacker goal. Table~\ref{resiliency_table} shows the resiliency table for our synthetic data, which conveys that an attacker cannot misclassify a normal patient into a High Cholesterol patient if he does not have access to more than two devices which refers that the system is 2-resilient for this specific goal. An attacker with the intent to misclassifying an excessive sweating state patient into a normal one or a high blood sugar patient into an abnormal oxygen level patient can become successful if he has access over one particular sensor. Similarly, for the UQVS dataset, it appears that changing a patient label from normal to decrease in hemoglobin oxygen saturation (DESAT) label is 20-resilient, which signifies, an attacker can not achieve this attack goal compromising 20 or fewer sensor measurements. Resiliency analysis capability of the framework provides a design guide specifying the relationship between the number of protected sensors with the reduction of risk.

\subsubsection{Scalability of SHChecker} \label{scalability}
We evaluate the SHChecker's scalability by analyzing the time requirement varying size of the SHS.Scalability of the model is mainly dependent on the time required to perform threat analysis for the solver based on attacker's capability and number of sensor measurements and this time is the most significant determinant of attack feasibility. We vary the number of sensors to construct our model for analyzing scalability of our system. Figures~\ref{fig:scalability}(a) and ~\ref{fig:scalability_real}(a) establishes that execution time to create clusters from the DBSCAN constraints take less time than the boundary creation time and this time increases linearly based on the number of sensor measurements. Figures~\ref{fig:scalability}(b) and \ref{fig:scalability_real}(b) infers that the construction of DBSCAN constraints requires a lot more time than that of DT clusters. However, as cluster, boundary creation and constraint generation are supposed to be performed beforehand, corresponding time requirements are insignificant for attack implementation. From Fig.~\ref{fig:scalability}(c), \ref{fig:scalability_real}(c), Fig.~\ref{fig:scalability}(d) and \ref{fig:scalability_real}(d), it is apparent that increasing the attacker's capability increases the execution time for the solver as it requires more constraints to check. The growth rate of time required for the solver performing real-time threat analysis corresponds to an exponential increment and raises scalability issues for large SHSs.

\begin{table}[t]
\centering
\caption{Complexity Analysis of DCMs based on the Number of Sensor Measurements} 
\small
\label{tab:clauses_dcm}
\begin{tabular}{|p{1.3cm}|c|c|c|c|}
\hline
\textbf{Dataset} & {\textbf{\begin{tabular}[c]{@{}c@{}}\# Measurements\end{tabular}}} &
\multicolumn{3}{c|}{\textbf{Number of Clauses}} \\
\cline{3-5}
  \multicolumn{1}{|c|}{} &
  \multicolumn{1}{|c|}{} &
  
  \multicolumn{1}{c|}{\textbf{NN}} &
  \multicolumn{1}{c|}{\textbf{DT}} &
  \multicolumn{1}{c|}{\textbf{LR}} \\ \hline
 & 8  & { 2652} & { 5196} & { 2148} \\
 & 10 & { 2694} & { 5104} & { 2160}  \\
 & 12 & { 2736} & { 5349} & { 2172}  \\
 & 14 & { 2778} & { 5465} & { 2184}  \\
\multirow{-4}{*}{\textbf{\begin{tabular}[c]{@{}c@{}}Synthetic\\  Data\end{tabular}}} &
  16 &
  { 2820} &
  { 897} &
  { 2196} \\ \hline
 & 26  & { 178962} & { 56813} & { 172608} \\
 & 27 & { 179005}  & { 56689} & { 172666} \\
 & 28    & {179048} & { 57071} & { 172724} \\
& 29    & { 179091} & { 57094} & { 172782} \\
\multirow{-4}{*}{\textbf{\begin{tabular}[c]{@{}c@{}}UQVS\\ Data\end{tabular}}} &
  30 &
  { 179134} &
  { 59124} &
  { 172840} \\ \hline
\end{tabular}
\end{table}

\begin{figure}[!t]
    \begin{center}
        \subfigure[]
        {
        \label{clauses_synthetic}
            \includegraphics[width=0.46\columnwidth]{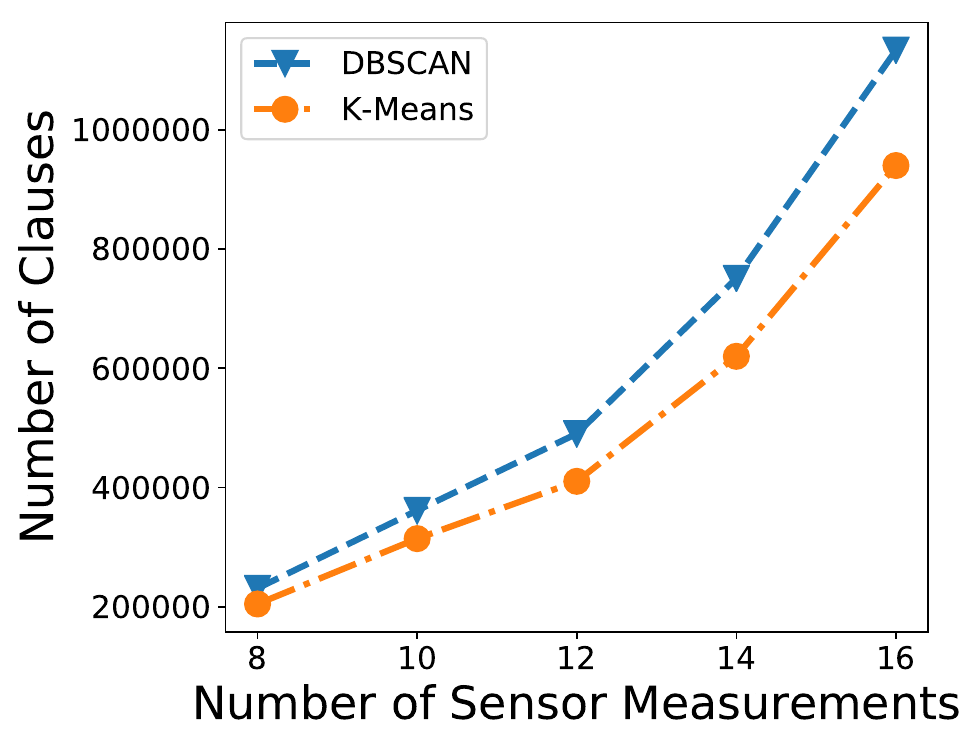}
        }
        \subfigure[]
         {
        \label{clauses_real}
            \includegraphics[width=0.46\columnwidth]{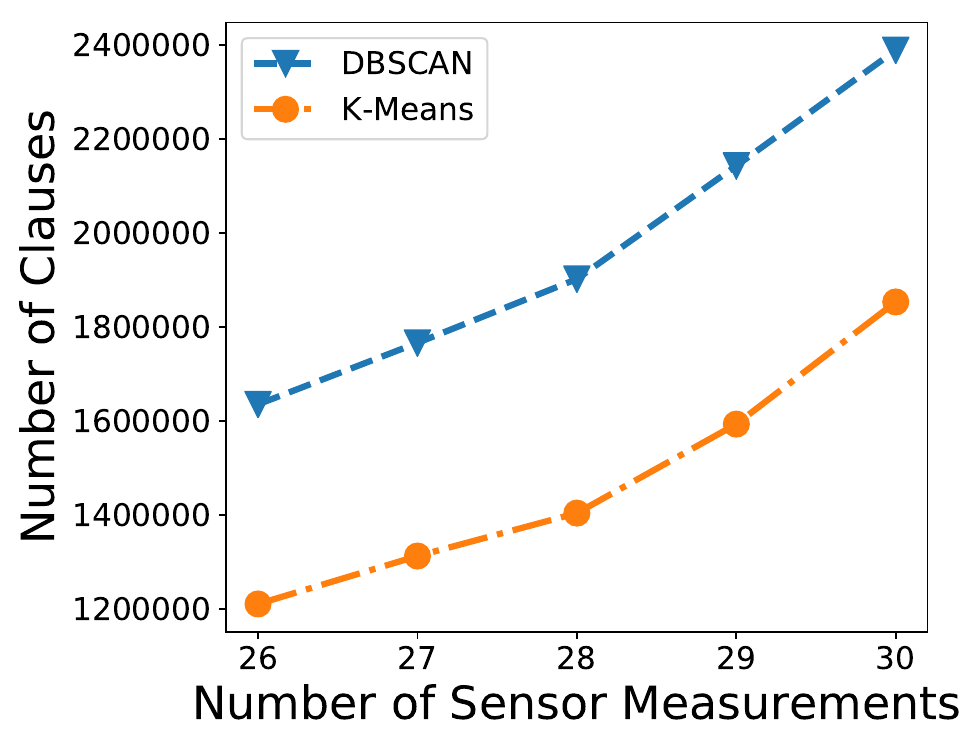}
        }
    \end{center}
    \caption{Complexity analysis of ADMs w.r.t. the number of measurements for (a) synthetic data and (b) UQVS data.}
    \label{fig:clauses_adm}
\end{figure}

{The complexity of the solver depends on the number of clauses. The time complexity is analyzed considering DT-based DCM and DBSCAN-based ADM. Table~\ref{tab:clauses_dcm} demonstrates the number of clauses for various DCMs with varied number of sensor measurements. It is apparent from the table that DCMs have limited dependency on the number of sensor measurements and all three have an almost similar number of clauses. For the NN model, the number of clauses depends on the size of the architecture. We used a 5-layer NN with 42, 44, 48 nodes in hidden layers for the real dataset. For the synthetic dataset, the number of nodes in the hidden layers was 20, 12, and 8. From Fig.~\ref{fig:clauses_adm}, we can observe that ADMs are mainly responsible for solver complexity as increasing the number of features adds a lot of clauses to the solver. However, it is also clear from the figure that both of the ADMs in consideration has given rise to a similar number of clauses. As a result, scalability analysis of other DT and DBSCAN is sufficient to understand the time requirements for other models. As the cluster, boundary creation, and constraint acquisition are performed beforehand, the feasibility of the attack is dependent on the time required to threat analysis only. From our experimentation, we found this value to be slightly more than 5 minutes for 30 sensor measurements. Hence, it is possible to launch an attack for a patient monitored by an SHS whose sensor measurements are not varying drastically with time.}



\subsubsection{Threat Analysis of Various ML Models} \label{Sec:threat_diff_ml}
\begin{figure}[!t]
    \begin{center}
        \subfigure[]
        {
        \label{threat_bar_chart}
            \includegraphics[width=0.46\columnwidth]{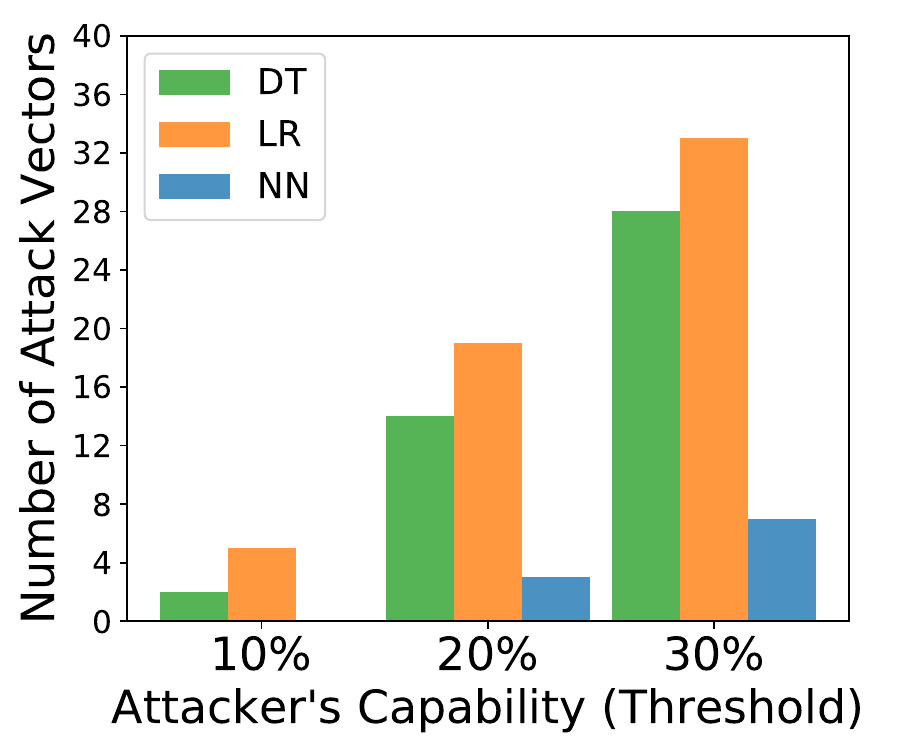}
        }
        \subfigure[]
         {
        \label{real_threat_bar_chart}
          \includegraphics[width=0.46\columnwidth]{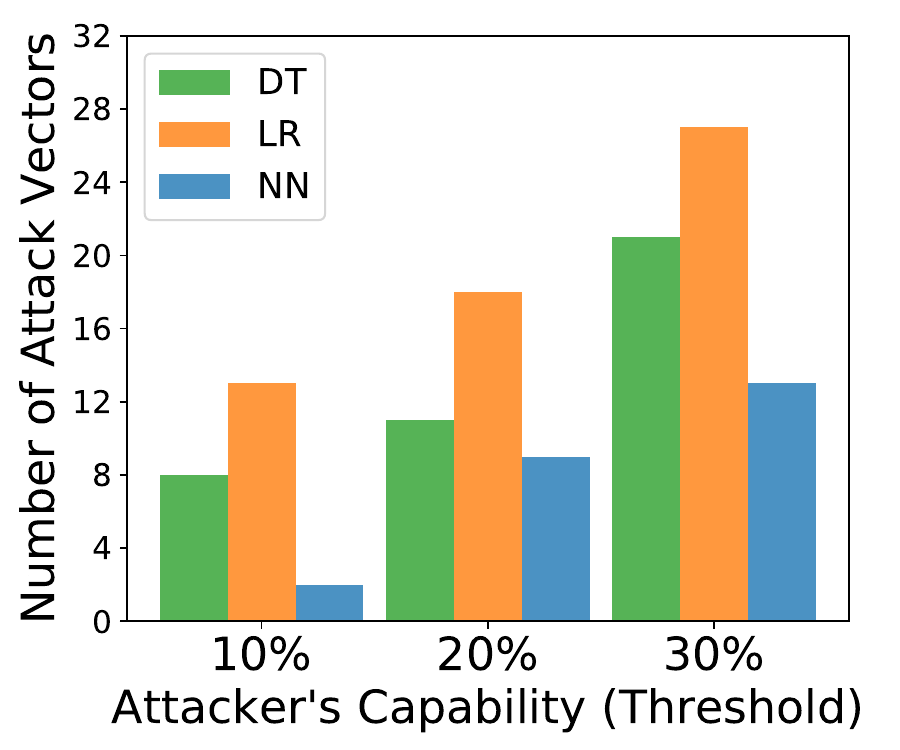}
        }
    \end{center}
    \caption{Threat analysis of various ML models for (a) synthetic data and (b) UQVS data}
    \label{fig:threat_all_ml}
\end{figure}

In this section, we analyze attack vectors of various ML models. Fig.~\ref{fig:threat_all_ml} shows the attack vector comparison varying DCMs for both of our dataset in consideration. 
{In this comparison, we considered DBSCAN as ADM.} Our comprehensive analysis shows that logistic regression-based DCM seems to be more vulnerable than the others. Besides, if the underlying DCM in use was NN instead of DT, the number of threat vectors would be fewer in number. The experimentation can be helpful while designing the decision control system. 
{Although the NN-based model's performance based on accuracy, precision, and recall is slightly less than the DT-based model, the DT-based model is subjected to more threats. Compromising an insignificant performance degradation can, therefore, increase the system's robustness against adversarial attacks. Hence, SHChecker can be proven helpful in providing a design guide to SHS design by assessing the threat of various ML models.}

\label{sec:evaluation}

\label{Sec:related_works}
\section{Related Work}
\label{Sec:related_works}
Due to vast advancements in medical and computer technologies, the healthcare system has experienced a growing interest from academia and industry. As a result, many works have been conducted to propose a secured IoMT-enabled SHS framework. A group of researchers proposes a cloud-enabled SHS framework deployed in an IoMT network~\cite{demirkan2013smart,catarinucci2015iot,panchatcharam2019internet}. Security is always a considerable concern while implementing a CPS. Popular cryptographic algorithms cannot be deployed on IoMT sensors due to limited computation power, which is one of the most alarming problems. Some research attempts to introduce a new cryptographic algorithm that does not require massive computational power. Gong et al. propose a  lightweight private homomorphism algorithm and an encryption algorithm developed using the concept of DES that can be implemented in the SHS \cite{gong2015medical}. Sharma et al. propose a privacy preservation scheme for wireless sensor network-based healthcare using multipath routing, secret sharing, and hashing \cite{sharma2018privacy}.   

Very few efforts have been made to assess threats in the SHS. Abie et al. propose a risk-based adaptive security framework for evaluating risk damages of eHealth and future benefits using the concept of game theory and context-awareness techniques as well \cite{abie2012risk}. He et at. demonstrate a high-level overview of existing security threats in SHS and presents a password security evaluation technique considering security vulnerabilities of password
building \cite{he2018privacy}. Newaz et al. performed adversarial machine learning-based threat analysis for SHS ~\cite{ newaz2020adversarial}. In their another works, they offered security solution for SHS by proposing machine learning-based IDS~\cite{newaz2019healthguard,newaz2020heka}. Darwish et al. present a mechanism for prioritizing Medical Internet of Things threats and aspects of the analysis that are likely to be affected by adding new devices in the system.\cite{darwish2017towards}. However, none of these work can evaluate all possible attack vectors based on the relationship between the patients' vital signs for threat analysis like us.

{Formal analysis of Deep Neural Network-based Machine learning models has gained a lot of focus in several contemporary researches. Several efficient tools (e.g., Reluplex,  Sherlock, Marabou, etc.) have been developed for verifying DNN~\cite{katz2017reluplex, dutta2017output, katz2019marabou}. Xiang et al. surveyed various state-of-the-art NN verification tools and presented and comprehensive comparison among them~\cite{xiang2018verification}. Verification of other ML-algorithms using formal modeling has been attempted. T$\ddot{o}$rnblom et al. presented a tool naming VoTE (Verifier of Tree Ensembles) for verifying DT based ensemble techniques that support up to 25 trees. Souri et al. formally verified a hybrid ML-based approach for fault prediction in IoT application that incorporates Multi-layer Perceptron (MLP) and Particle Swarm Optimization algorithms~\cite{souri2020formal}. Unlike these research efforts, our proposed framework performs formal threat analysis of black-box SHS incorporating two different purpose ML-based models and has opened a novel research direction in the ML-based formal modeling domain.}

\label{Sec:conclusion}

\balance

\section{Conclusion}
\label{Sec:conclusion}
In this paper, we present SHChecker, a machine learning and formal modeling-based framework, to model and study the security of SHSs. The tool can analyze the potential threats that satisfy the attacker's goal. We use machine learning to understand the relationships between the sensor measurements, health states, and their consistency. We exploit this knowledge to perform formal analysis to synthesize potential attack vectors for a given attack model, where the attacker can change the health state (the actual one) to a wrong one (the targeted state). Our evaluation results on both of our datasets in consideration show that diverse attacks are possible by compromising different numbers and types of sensor values, even compromising only one sensor measurement. Though our current model considers only a single label attack, we plan to expand it to the multi-label goals in our future work. Besides, we plan to study other efficient machine learning algorithms to improve the performance of our proposed SHChecker framework. As a pair of feature considerations in our anomaly detection model can still susceptible to some false negative outcome, we aim to create formal model constraints considering all feature relations together in our future work.

\bibliographystyle{unsrt}
\bibliography{References} 

\end{document}